\documentclass{jpp}
\usepackage{graphicx}

\usepackage[utf8]{inputenc}
\usepackage[T1]{fontenc}
\usepackage{amsmath}
\usepackage{cancel}
\usepackage[dvipsnames]{xcolor}
\usepackage{xcolor} 
\usepackage{enumerate}
\shorttitle{RMF plugging}
\shortauthor{Miller, Be'ery, Gudinetsky, and Barth}

\title{Plugging of multi-mirror machines by a traveling rotating magnetic field}

\author{Tal Miller\aff{1},  Eli Gudinetsky\aff{1}, Ilan Be'ery\aff{2}, Ido Barth\aff{1}
 \corresp{\email{ido.barth@mail.huji.ac.il}}}
\affiliation{\aff{1} Racah Institute of Physics, The Hebrew University of Jerusalem, Jerusalem, 91904 Israel
\aff{2} nT-Tao, 5 Ha-Nagar st., Ramat Hasharon, 4526005 Israel}

\newcommand{\ie}{i.e., }

\begin{document}

\maketitle

\date{\today}

\begin{abstract}

Axial plugging is a critical challenge for fusion in open-ended magnetic confinement systems.
Unlike simple magnetic mirrors, which suffer from direct axial flow, multi-mirror systems utilize a series of aligned magnetic cells to suppress plasma loss; however, the resulting confinement still requires additional plugging to reach Lawson criterion levels.
In [T. Miller et al., Phys. Plasmas 30, 072510 (2023)], it was found that applying a traveling and rotating electric field in multi-mirror machines can significantly suppress axial loss due to a selectivity effect induced by the Doppler shift of the ion cyclotron resonance.
However, this method is energetically expensive and vulnerable to plasma screening effects.
Here, we show that using a traveling, rotating magnetic field can achieve comparable plugging effectiveness while offering better penetration and lower energy costs. 
Two limiting scenarios, with and without an induced electric field, were considered.
The confinement enhancement is calculated using a semi-kinetic rate equation model, in which the rate coefficients are determined from single-particle simulations.
While both scenarios exhibit significant confinement enhancement, the scenario without an induced electric field is much more energetically efficient, as it relies on phase-space mixing rather than on energy deposition in the escaping particles.
The decoupling of confinement from plasma collisionality enables fusion conditions in the central cell while allowing affordable and efficient confinement enhancement in the multi-mirror sections. 
\end{abstract}
\section{Introduction}
\label{sec: intro}

Multi-mirrors (MM) are linear magnetically confined systems composed of a main, fusion-aimed, cell and two sections of multiple magnetic mirrors attached on each side of the system to reduce axial loss through the loss cone [\cite{post1967confinement, logan1972multiple, logan1972experimental, mirnov1972gas, makhijani1974plasma, tuszewski1977transient, burdakov2016multiple, budker1971influence, mirnov1996multiple, kotelnikov2007new, mirnov1972gas}].
The confinement enhancement in simple MM systems is based on collisional scattering of particles inside the adjacent mirror cells. 
In this picture, particles escaping the central cell undergo multiple scattering events in the MM sections, effectively increasing their residence time in the system. 
The efficiency of this mechanism depends strongly on plasma collisionality, where higher collisionality leads to greater redistribution of particle trajectories and stronger confinement, but at the cost of requiring lower temperatures and higher densities that are less favorable for fusion performance  [\cite{post1967confinement, kotelnikov2007new, miller2021rate}].

Other methods to increase axial confinement in linear machines include tandem mirror machines with thermal barriers [\cite{inutake1985thermal, grubb1984thermal, katanuma1986thermal, pratt2006global, tamano1995tandem, ivanov2013gas, ivanov2017gas, anderson2020introducing, forest2022physics, egedal2022fusion, endrizzi2023}] and additional radio-frequency (RF) plugs [\cite{golovato1985fueling}], diamagnetic confinement [\cite{beklemishev2016diamagnetic, kotelnikov2020structure}] multi-mirrors (MM) systems [\cite{post1967confinement, logan1972multiple, logan1972experimental, mirnov1972gas, makhijani1974plasma, tuszewski1977transient, burdakov2016multiple, budker1971influence, mirnov1996multiple, kotelnikov2007new, mirnov1972gas}], asymmetric MM [\cite{post1981particle}], moving multi-mirrors [\cite{tuck1968reduction, budker1982gas}], helical mirror with rotating plasma [\cite{beklemishev2013helicoidal, postupaev2016helical, sudnikov2019first, ivanov2021long, sudnikov2022plasma,tolkachev2024electromagnetic}], ponderomotive RF plugging [\cite{motz1967radio, watari1974theory, hatori1975critical, watari1978radio, uehara1978radio, hiroe1978experiment, fader1981rf, fisch2003current, dodin2004ponderomotive, dodin2005ponderomotive}], and side-sections of field-reversed configuration (FRC) at the mirror throats [\cite{shi2019magnetic}].

In MM systems, the escaping axial flux scales inversely with the number of cells in the MM section, where the scaling law of the outgoing flux with system size depends on the underlying thermodynamic scenario.
The most favorable confinement arises in isentropic systems, where adiabatic cooling reduces the density and shortens the mean free path (MFP) along the MM section [\cite{miller2021rate}].
Nevertheless, even in this optimistic scenario, the confinement time scales with the system length such that an impractically large number of cells would be required to meet the Lawson criterion [\cite{lawson1957some, wurzel2022progress}].
This result highlights the need to develop confinement-enhancement strategies for achieving sustainable fusion in MM devices.
One idea was to exploit the directionality of the outgoing flux in MM systems and apply a drive that affects incoming and outgoing particles differently. 
The general concept of asymmetric drive was introduced and analyzed in [\cite{post1981particle}].

In a previous study, we proposed such an asymmetric drive by applying an external traveling rotating electric field (TREF) resonantly coupled with the ion cyclotron frequency in the frame of reference of the outgoing particles [\cite{miller2023rf}]. 
More specifically, we employed a radially rotating electric field with a frequency slightly detuned from the ions' exact Larmor frequency and with a non-zero axial wave vector.
The Doppler shift of such an RF configuration compensates for the resonance mismatch of outgoing particles solely, while mildly affecting returning particles with opposite axial velocities.
This selection or asymmetric effect yields a significant confinement enhancement effect, beyond the stationary RF plugging effect considered in Tandem machines [\cite{golovato1985fueling}].
The analysis of the TREF effect in MM systems was studied using single-particle simulations and a revised semi-kinetic rate equation model, in which the ions in each mirror cell are divided into three populations (confined, escaping, and returning) and includes the processes of Coulomb scattering within each cell, the thermal transmission between neighboring cells, and RF-induced transition rates deduced from Monte-Carlo single-particle simulations.
The steady-state axial density profile and the outgoing flux are then calculated from the rate equations' stationary solution [\cite{miller2021rate, miller2023rf}].
Yet, since the re-confinement mechanism of TREF is based on the injection of perpendicular energy to loss cone particles, the energy cost is not negligible and might even make TREF energetically unprofitable.
Remarkably, an increase in perpendicular energy is not, in itself, a necessary condition for particle re-trapping.
Instead, it is sufficient to induce energy mixing between the axial and perpendicular motions, while preserving the total energy, as in elastic collisions.
Accordingly, it would be advantageous to develop a mechanism that reduces the effective MFP in the MM sections, while leaving the Coulomb MFP in the central cell, where the plasma remains in an efficient fusion regime (MFP $\approx$1~km) unchanged.
Interestingly, a resonant magnetic field may fulfill these requirements.

In this work, we propose using a traveling rotating magnetic field (TRMF) instead of TREF to enhance the axial confinement in MM systems.
We demonstrate that, while TREF is based on injecting transverse energy into the particles, TRMF induces phase space mixing without the need to heat the ions, as the magnetic force does not perform work on the charged particles.
Still, TRMF exhibits plugging efficiency similar to TREF while consuming little energy. 
As with TREF, the resonant coupling arises from the asymmetry induced by the Doppler-compensated field phase velocity.
An additional advantage is that transverse magnetic fields penetrate plasmas more easily than electric fields [\cite{watari1978radio, hugrass1981numerical, milroy1999numerical}], making TRMF more feasible than TREF for dense plasma conditions. 
This capability of a rotating magnetic field (RMF) is well established in the context of driving and sustaining FRCs [\cite{jones1999review, furukawa2019verification, polzin2020state, cohen2023laboratory}], providing strong evidence for efficient magnetic-field penetration into the plasma.
In contrast, the axial electric fields induced by the TRMF will decay in the plasma, thereby reducing plasma heating in the MM section that does not contribute to fusion or to axial re-trapping.
Since the plugging effect is robust to the degree of decay of the induced axial electric field, it is effective both in dense and dilute plasmas.

The structure of the paper is as follows.
Sec.~\ref{sec: fields} introduces the configuration of the static magnetic mirror field and the external TRMF.
Sec.~\ref{sec: single particle} studies the particle dynamics in the presence of TRMF using Monte-Carlo single particle simulations.
The simulation results are then integrated over the fuel-particle distribution to evaluate the transition rates between different phase-space populations, and are compared with those obtained for the TREF scheme.
Sec.~\ref{sec: rate eqs} incorporates the effect of TRMF into the generalized rate equations model for MM systems and calculates the confinement enhancement.
Sec.~\ref{sec: discussion} discusses the power considerations and the trapping mechanism of the different RF schemes.
Finally, Sec.~\ref{sec: conclusions} summarizes the conclusions.
\section{Fields configuration}
\label{sec: fields}

\begin{figure}
    \centering
    \includegraphics[clip, trim=0.0cm 0.0cm 1.cm 0.0cm, width=0.7\linewidth]{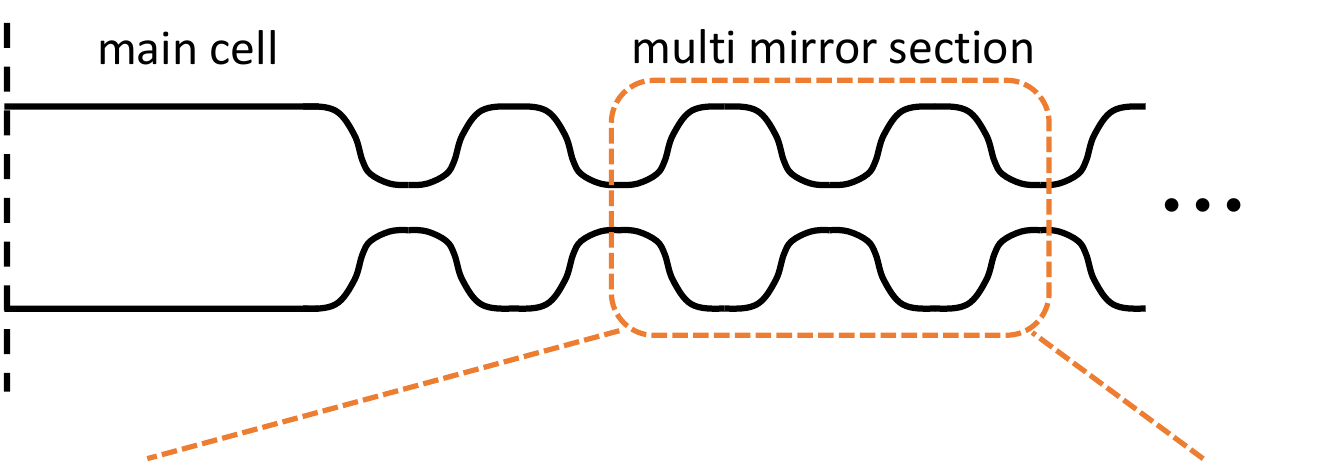}
    \includegraphics[clip, trim=.5cm 0.4cm .0cm 0.4cm, width=0.7\linewidth]{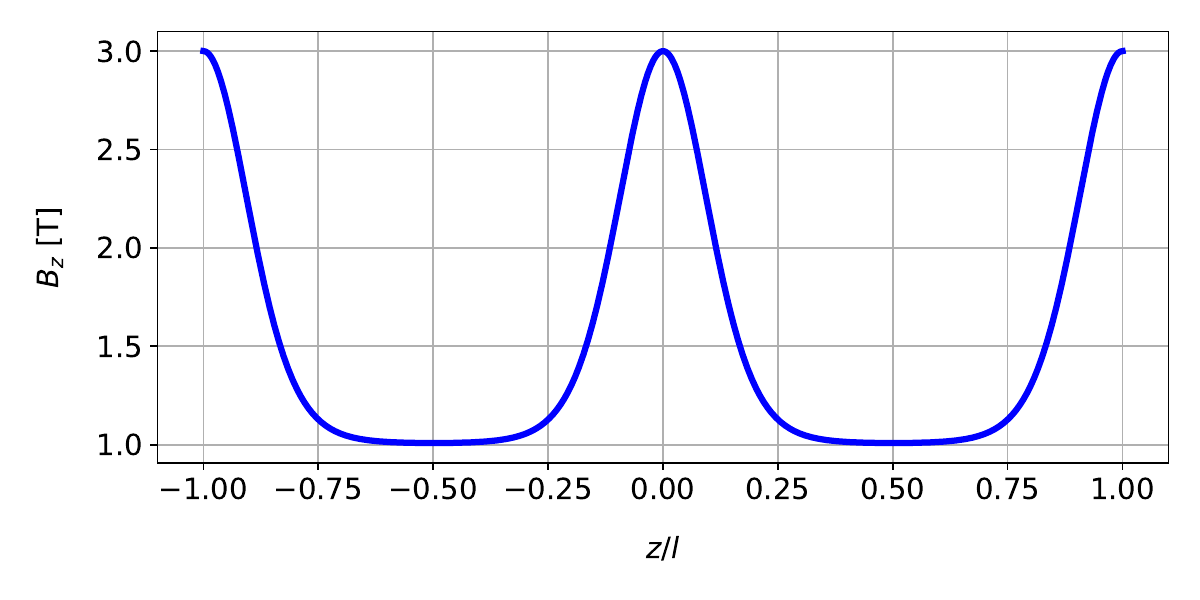}
    \caption{An illustration of one section of an MM system (top) and the amplitude of the axial magnetic field (Eq.~(\ref{eq: Bz mirror})) of two MM cells with $R_m=3$ (bottom).
    The illustration shows the right half of the MM system, where the left half is assumed to lie to the left of the vertical dashed line.}
    \label{fig: axial magnetic field profile}
\end{figure}

The classic MM system comprises one central fusion cell and two MM sections [\cite{post1967confinement, logan1972multiple, logan1972experimental, mirnov1972gas, makhijani1974plasma, tuszewski1977transient, burdakov2016multiple, budker1971influence, mirnov1996multiple, kotelnikov2007new, mirnov1972gas}].
Following [\cite{post1967confinement}], we model the MM section by a periodic magnetic field with an axial component of the form
\begin{eqnarray}
    \label{eq: Bz mirror}
    B_{z} = B_{0}\left[1+\left(R_{m}-1\right)\exp\left(-5.5\sin^{2}\frac{\pi z}{l}\right)\right]
\end{eqnarray}

\noindent where $B_0$ is the minimal magnetic field,
$R_m = B_{max} / B_0$ is the mirror ratio, and $l$ is the length of each MM cell.
Fig.~\ref{fig: axial magnetic field profile} illustrates the MM system and the axial component of the magnetic field, $B_z$.
To satisfy Gauss's law $\nabla\cdot\mathbf{B}=0$, we include $x,y$ components in the MM magnetic field such that the total (static) field near the mirror axis reads
\begin{eqnarray}
    \mathbf{B}_{\text{mirror}}=-\frac{1}{2}\frac{\partial B_{z}}{\partial z}\left(x\hat{x}+y\hat{y}\right)+B_{z}\hat{z}
    \label{eq: B mirror}
\end{eqnarray}

As discussed in the introduction, we consider time-dependent TRMF and examine the effect of phase-space mixing and the resulting enhanced confinement with minimal plasma heating.
First, a simple rotating magnetic field (RMF) in the mirror transverse plane ($x,y$) but at a fixed axial $(z)$ location, can be written in phasor form as
\begin{eqnarray} \label{B_RMF}
\mathbf{B}_{\mathrm{RMF}}&=B_{1}e^{-i\omega t} \left(\hat{x}-i\hat{y}\right),
\end{eqnarray}
where $B_{1}$ is the field amplitude, $\omega$ is the RF angular frequency of the rotating field, and the rotation direction is the same as that of the cyclotron motion of positively charged ions for $B_0>0$.
The corresponding induced electric field, in this case, can be approximated to first order in $\omega r/c$, with $r$ the distance from the mirror axis (in cgs convention) as
\begin{eqnarray}
\mathbf{E}_{\mathrm{RMF}}&=-B_{1}\omega e^{-i\omega t} \left(x-iy\right)\hat{z}.
\end{eqnarray}
It is noted that the induced electrostatic field vanishes on the mirror axis and grows linearly with $r$. 

To extend the magnetic field to a traveling field along the axial $\hat{z}$ direction (\ie a TRMF), we define the field's phase as
\begin{eqnarray}
    \varphi \left(z, t \right)=k z-\omega t.
\end{eqnarray}
In this case, both the magnetic and the induced electric fields should include spatial corrections, which, to the first order, read
\begin{eqnarray}\label{eq:TRMF_B}
\mathbf{B}_{\mathrm{TRMF}}&=&B_{1}e^{i\varphi}\left[\left(\hat{x}-i\hat{y}\right)+k\left(y+ix\right)\hat{z}\right] \\
\mathbf{E}_{\mathrm{TRMF}}&=&B_{1}\omega e^{i\varphi}\left[-\left(x-iy\right)\hat{z} + kxy\left(\hat{x}+i\hat{y}\right)\right]
\label{eq:TRMF_E}
\end{eqnarray}
where the two small parameters in the approximation are $\omega r/c \ll 1$ and  $kr \ll 1$ (see Appendix).
This expression for the induced electric field assumes vacuum conditions, whereas in plasmas, the electric field may be suppressed or penetrate less effectively than the magnetic field.
Thus, we also analyze the case in which it is absent, realized by setting it to zero in part of the single-particle calculations in Sec.~\ref{sec: single particle}. 
Notably, in this scenario, the TRMF does not transfer energy to the particle and therefore does not contribute to plasma heating.
Hereafter, we refer to this scheme as TRMF--noE.

For completeness and comparison, we will also consider and simulate the TREF plugging method that was first proposed in [\cite{miller2023rf}].
The expressions (in cgs convention) for the electric and the (induced) magnetic fields in the TREF method, which are analogous to those of the TRMF method, but with a role-exchange of the magnetic and electric fields, read
\begin{eqnarray} 
\label{eq:TREF_E}
\mathbf{E}_{\mathrm{TREF}}&=&E_1e^{i\varphi}\left[\left(\hat{x}-i\hat{y}\right)+k\left(y+ix\right)\hat{z}\right] \\
\mathbf{B}_{\mathrm{TREF}}&=&E_1\frac{\omega}{c^2} e^{i\varphi}\left[\left(x-iy\right)\hat{z} - kxy\left(\hat{x}+i\hat{y}\right)\right],
\label{eq:TREF_B}
\end{eqnarray}
where $E_1$ is the TREF electric field amplitude.
Here, we also include the first-order temporal ($\omega r/c$) and spatial ($kr$) correction terms for both electric and induced magnetic fields, which were neglected in [\cite{miller2023rf}].
Nonetheless, the forthcoming simulation results show that this assumption is justified for TREF.

In the next section, we consider three scenarios: TRMF with and without an induced electric field, and TREF with an induced magnetic field, although for the parameters considered, its effect on the dynamics is negligible.
The two TRMF scenarios, \ie with and without an induced electric field, can serve as models for the limiting behavior of dilute laboratory plasmas (where the electric field can penetrate readily) and for dense fusion plasmas (where the electric field is strongly suppressed). 
A crucial caveat is that, in the high-density plasma limit, the magnetic field itself might be suppressed or modified by the plasma, thereby affecting the efficiency of TRMF plugging. 
A detailed study of this collective effect is left for future work.
\section{Single particle simulations}
\label{sec: single particle}

To quantify the RF effect on particles in the mirror system, we employ the single-particle approximation and perform a Monte-Carlo analysis, where the initial velocities were sampled from a thermal distribution with an ion temperature of $k_{B} T_i=10~\mathrm{keV}$ and a uniformly distributed random direction. 
The ions were initialized at the mirror midplane ($z = l/2$), with radial positions, $r_0$, sampled uniformly from the range $[0, 10]~\mathrm{cm}$ to represent a realistic plasma profile.
The azimuthal angle (in the $x$-$y$ plane) was also sampled uniformly over the interval $[0, 2\pi]$ to simulate a random phase relative to the phase of the RF field.
The static mirroring magnetic field modeled one MM cell with axial magnetic field as defined in Eq.~(\ref{eq: Bz mirror}), where $l=1~\mathrm{m}$, $B_0=1~\mathrm{T}$, and $R_m=5$. 
For the time-dependent RF fields, we considered 3 cases of RF fields, as described in Sec.~\ref {sec: fields}: 
(a) TRMF [Eqs.~(\ref{eq:TRMF_B}), and (\ref{eq:TRMF_E})], 
(b) TRMF--noE  [Eq.~(\ref{eq:TRMF_B}) solely],
and (c) TREF [Eqs.~(\ref{eq:TREF_E}) and (\ref{eq:TREF_B})].

For TRMF, \ie cases (a) and (b), the magnetic field amplitude was $B_{RF}=0.05~\text{T}$ while for TREF of case (c), the electric field amplitude was $E_{RF}=50~\text{kV/m}$.
In case (c), we found that the effect of the induced magnetic field [Eq.~(\ref{eq:TREF_B})] is negligible, so we do not show a case of TREF without the induced magnetic field.
Neglecting collisions and collective effects, we calculate the trajectories of deuterium and tritium ions under the influence of the Lorentz force $\mathbf{F}=q\left(\mathbf{E}+\mathbf{v}\times\mathbf{B}\right)$.
We employed a symplectic scheme, which preserves phase space volume [\cite{he2015volume}], to solve the particles' trajectories under the influence of the static and RF fields.
The simulation time for all particles of the same isotope was identical, chosen to be the thermal passage time of a particle along one MM cell length, $\tau_\text{th} \equiv l/v_\text{th}$, where $v_\text{th}$ is the ion's thermal velocity (different for deuterium and tritium ions). 

The ion cyclotron frequencies of deuterium and tritium at the mirror midplane are
$\omega_{0,D} = eB_0/m_D = 2\pi\times 7.6$~MHz and $\omega_{0,T} = eB_0/m_T = 2\pi\times 5$~MHz, respectively. 
Note that in the figures, the RF frequency $\omega$ is usually normalized by $\omega_{0,T}$.
The time step in the simulations was set to be $\Delta t=\tau _{cyc}/50$ where $\tau _{cyc}=2\pi / \omega_{0D,0T}$ is the cyclotron period for the relevant isotope of each calculation (tested for numerical convergence).
Also note that although the calculations were performed separately for deuterium and tritium, both present in a D–T fusion reactor but exhibiting distinct dynamics due to their different masses, we show only the tritium results in some of the figures to avoid clutter. 
The key results related to the plugging enhancement, however, are presented for both isotopes.

For each RF scheme, we survey a range of RF parameters, $k$ and $\omega$, and for each configuration, we perform $10{,}000$ Monte Carlo realizations of the initial conditions to obtain reliable statistics on the kinetic transition rates.
The range selected for the survey in the $k$--$\omega$ parameter space was chosen to remain consistent with the validity limits of the first-order approximation in Eqs.~(\ref{eq:TRMF_B})-(\ref{eq:TRMF_E}), \ie $\omega r / c \ll 1$ and $k r \ll 1$,  and with the system's physical dimensions.
Consequently, the RF frequency $\omega$ was chosen in the range $\omega \in [0.5, 2]\,\omega_{0,T}$, roughly centered around the tritium cyclotron frequency, while the wave-vector range was $k \in 2\pi \times [-1,\, 1]~\mathrm{m}^{-1}$, 
corresponding to wavelengths bounded by the MM cell length.
The resulting restrictions on the particle position are $r\ll c/\omega_{max}=5$~m and $r \ll 1/k_{max}=16$~cm. 
These conditions are satisfied by the particles' initial radii in our simulations, which were set to be at most $10$~cm.

In the semi-kinetic rate equation model (see Sec.~\ref{sec: rate eqs} below), we classify the particles into three populations according to their initial conditions: (a) confined particles, (b) right-going particles, and (c) left-going particles. 
In our convention (see Fig.~\ref{fig: axial magnetic field profile}), right-going particles escape outward from the MM system, while left-going particles propagate inward toward the central fusion cell (the roles would be reversed for the left half of the MM system).
This classification is performed at the beginning of the simulation, by checking the loss cone condition at the midplane $\left( v_{\perp}/v \right)^2 < B_\text{min} / B_\text{max}$, where $v_{\perp}$ is the perpendicular velocity and $v$ is the total velocity.

We denote the number of particles in each population by $N_{c}$ for confined particles, $N_{r}$ for right-going (escaping), and $N_{l}$ for left-going (incoming) particles.
During the simulations, we take $50$ equidistant time snapshots and check, for each particle, the local loss cone condition $\left( v_{\perp}/v \right)^2 < B / B_\text{max}$, where $B$ is the local mirror magnetic field at the particle's location at a given time.
We then track the time evolution of each particle’s classification, which is updated whenever the particle crosses one of the loss-cone boundaries.
The critical parameter required to estimate the overall efficiency of RF plugging in MM machines via the rate equation model (see Sec.~\ref{sec: rate eqs}) is the number of "converted" particles, say, that originated at the right loss cone but ended up confined $\Delta N_{r \rightarrow c}$. 
Let us define the time-dependent population conversion metric,
\begin{equation} 
    \label{eq: delta_N_bar}
    \Delta \bar{N}_{ij} \left( t \right) 
     \equiv \frac{\Delta N_{i \rightarrow j}(t)}{N_i (t=0)},
    \qquad ij \in \{rc, cr, lc, cl, rl, lr\},
\end{equation}
for the six possible transitions. 
These quantities are normalized by the original number of particles in the source population, so by definition $\Delta \bar{N}_{ij} \left( t \right)\in [0,\, 1]$.

In Figs. \ref{fig: population conversion T, TRMF}-\ref{fig: population conversion T, TREF}, we plot the time evolution of the population conversion metric for tritium for different values of the RF-parameters $k$ and $\omega$, and for the different RF schemes, TRMF, TRMF--noE, and TREF, respectively.
Each subplot in the figures corresponds to a specific pair of $k$ and $\omega$ values, where the horizontal axis shows the time over the interval $[0,\, \tau_{\mathrm{th}}]$, and the vertical axis displays the population–conversion metric over the range $[0,\, 1]$.
The corresponding plots for deuterium are similar to those for tritium, so we omit them here.
The fluctuations observed in some of the trajectories arise from particles switching identities multiple times, oscillating between different populations.
\begin{figure}
    \centering
    \includegraphics[clip, trim=1.0cm 0.3cm 0.0cm 1.8cm, width=1.0\linewidth]{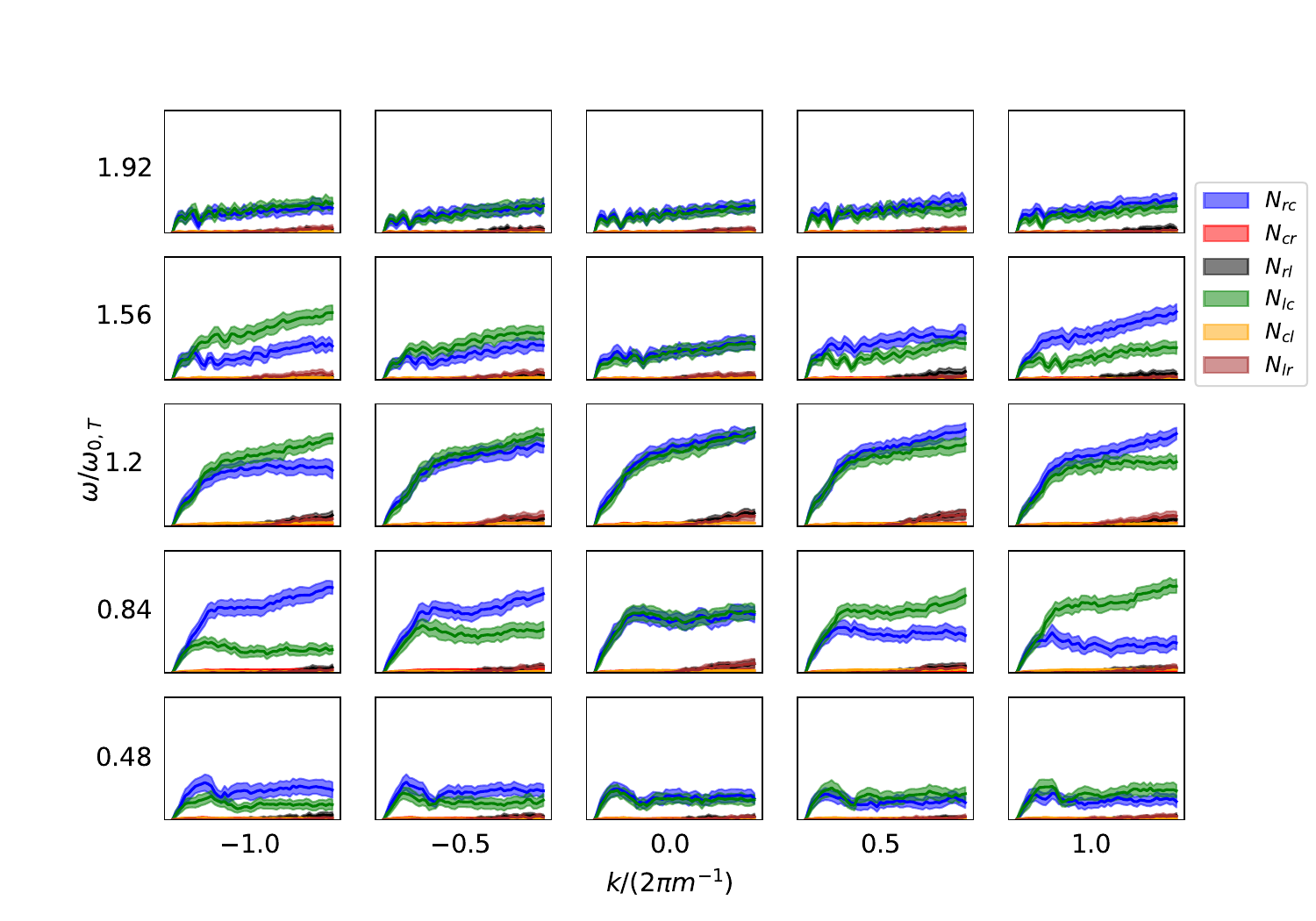}
    \caption{Population conversion $\Delta \bar{N}_{ij}$ plots for tritium in TRMF for different values of $k,\omega$.
    Colors indicate different transitions between the three populations (see legend).
    In each subplot, the horizontal axis represents time over the interval $[0,\, \tau_{\mathrm{th}}]$, while the vertical axis shows the population–conversion metric over the range $[0,\, 1]$.
    }
    \label{fig: population conversion T, TRMF}
\end{figure}
\begin{figure}
    \centering
    \includegraphics[clip, trim=1.0cm 0.3cm 0.0cm 1.8cm, width=1.0\linewidth]{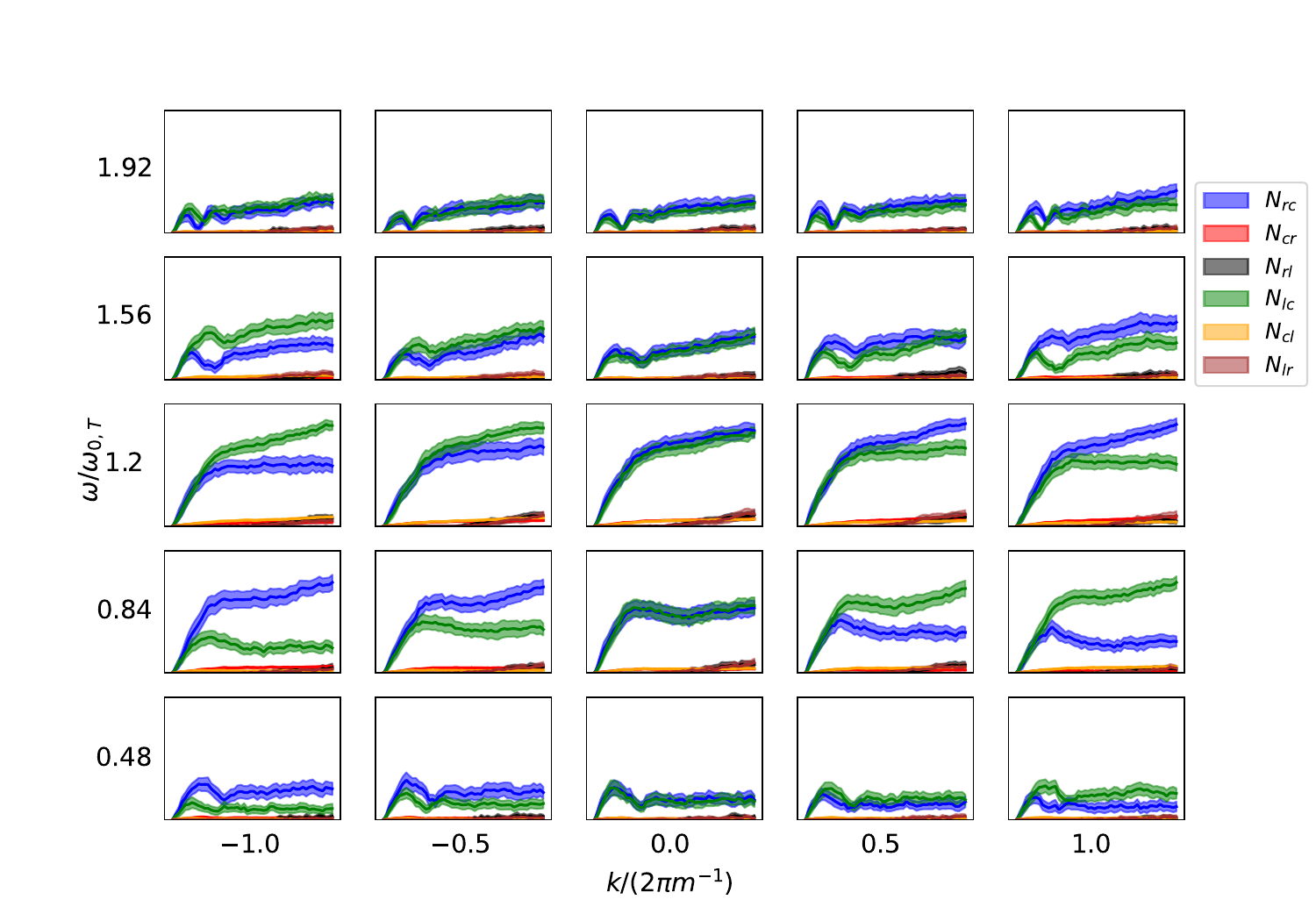}
    \caption{Population conversion $\Delta \bar{N}_{ij}$ plots for tritium in TRMF--noE for different values of $k,\omega$.
    Colors indicate different transitions between the three populations (see legend).
    In each subplot, the horizontal axis represents time over the interval $[0,\, \tau_{\mathrm{th}}]$, while the vertical axis shows the population–conversion metric over the range $[0,\, 1]$.}
    \label{fig: population conversion T, TRMF iff=0}
\end{figure}
\begin{figure}
    \centering
    \includegraphics[clip, trim=1.0cm 0.3cm 0.0cm 1.8cm, width=1.0\linewidth]{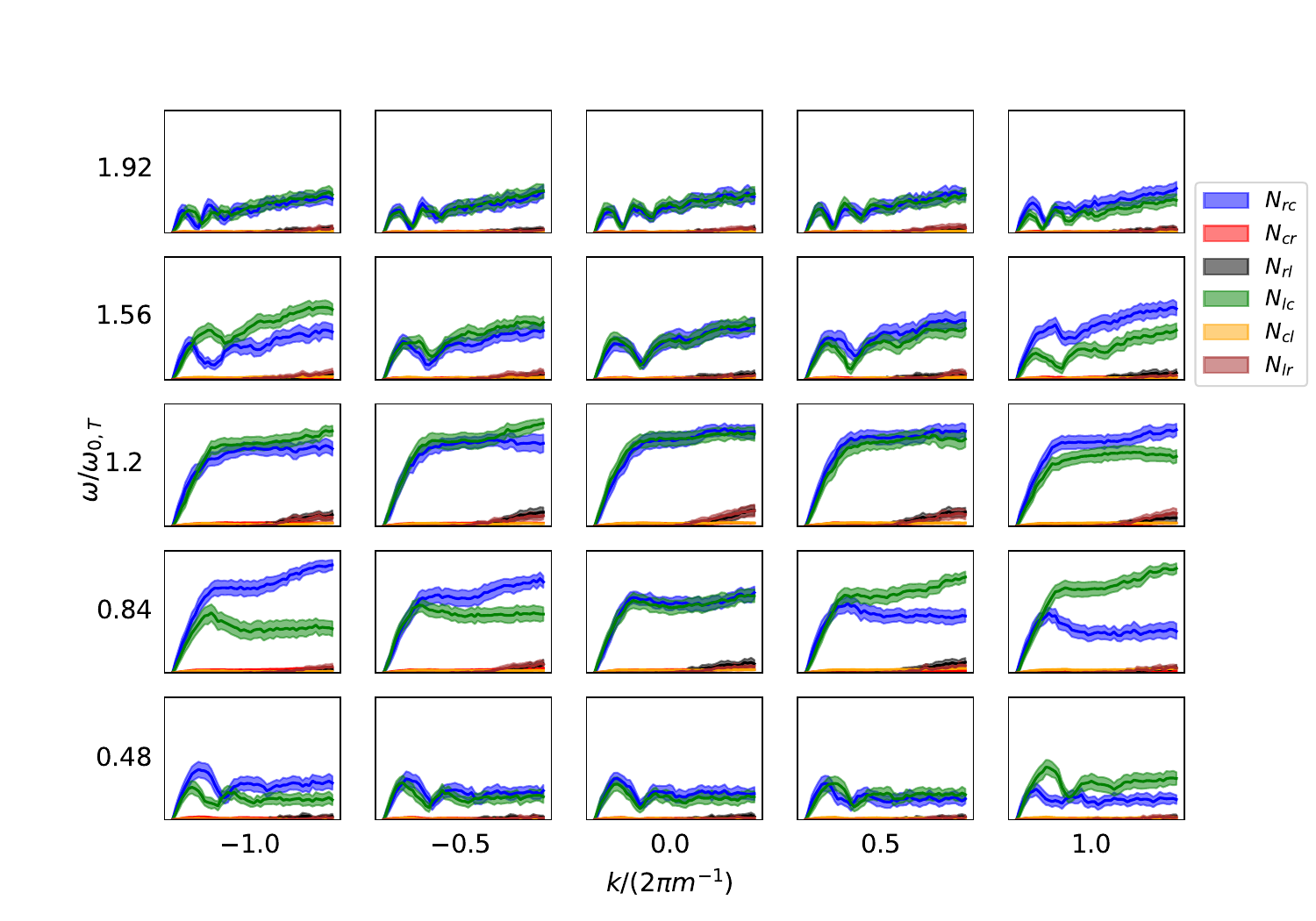}
    \caption{Population conversion $\Delta \bar{N}_{ij}$ plots for tritium in TREF for different values of $k,\omega$.
    Colors indicate different transitions between the three populations (see legend).
    In each subplot, the horizontal axis represents time over the interval $[0,\, \tau_{\mathrm{th}}]$, while the vertical axis shows the population–conversion metric over the range $[0,\, 1]$.}
    \label{fig: population conversion T, TREF}
\end{figure}
To estimate the statistical error associated with the finite number of Monte-Carlo particle samples, we apply a bootstrap procedure in which $50$ resampled subsets (with repetitions) are drawn from the original set of $10,000$ particles, and the population-conversion metric is recomputed for each subset. 
In the population-conversion plots, the solid lines denote the median, while the shaded regions indicate the $\pm 2 \sigma$ bounds at each time point.

For $k=0$, the RF fields are symmetric in the $\pm \hat{z}$ directions, so particles traveling in opposing axial directions are expected to behave similarly.
This expectation is consistent, within the numerical accuracy indicated by the shaded regions, with the results presented in Figs.~\ref{fig: population conversion T, TRMF}-\ref{fig: population conversion T, TREF}).
For example, this symmetry is evident in the comparison of $\Delta \bar{N}_{rc}$ (blue) and $\Delta \bar{N}_{lc}$ (green) in the middle column of the figures.
Due to symmetry, for a fixed $\omega$, the solutions corresponding to opposing wave vectors $\pm k$ are expected to be antisymmetric under the transformation $\hat{z} \rightarrow -\hat{z}$. 
This behavior is clearly visible in the curves in the panels with positive $k$ when compared to the corresponding panels with negative $k$.

The next step is to evaluate from the single-particle simulations the transition rates between the different populations, which will serve as the transition coefficients in the rate equation model to be introduced in Sec.~\ref{sec: rate eqs}.
To this end, we extract the population conversion values at the end of each simulation  $\bar{N}_{ij} \equiv \Delta \bar{N}_{ij} \left( \tau_{\mathrm{th}} \right)$, defined as the average over the final five time frames. 
To get RF-induced transition rates in units of inverse time, we divide by $\tau_\text{th}$, the characteristic timescale for an average non-trapped particle to travel a single mirror cell,
\begin{equation}
    \label{eq: RF rate}
    \nu_{RF,ij} \equiv  \frac{\bar{N}_{ij}}{\tau_\text{th}} =  \frac{ \Delta \bar{N}_{ij} \left( \tau_{\mathrm{th}} \right) }{\tau_\text{th}}.    
\end{equation}

For each RF scheme, we plot the RF-induced transition rates (six panels per figure) for tritium, obtained from the single-particle Monte Carlo simulations.
The schemes presented are TRMF (Fig.~\ref{fig: smoothed RF rates T, TRMF}), TRMF--noE (Fig.~\ref{fig: smoothed RF rates T, TRMF iff=0}), and TREF (Fig.~\ref{fig: smoothed RF rates T, TREF}).
For visualization purposes, we plot the dimensionless RF rates $ \bar{N}_{ij}= \nu_{RF,ij} \tau_\text{th}$ after applying a Gaussian filter to smooth the RF rate maps and reduce numerical noise.
For comparison, we also conducted a full set of Monte Carlo simulations to obtain the transition rates for deuterium, the second component of the D-T fuel.
The results are similar to those of tritium, and for illustration, we present them in Fig.~\ref{fig: smoothed RF rates D, TRMF iff=0} for the case of TRMF--noE.

\begin{figure}
    \centering
    \includegraphics[clip, trim=0.0cm 0.0cm 0.0cm 1.0cm, width=1.0\linewidth]{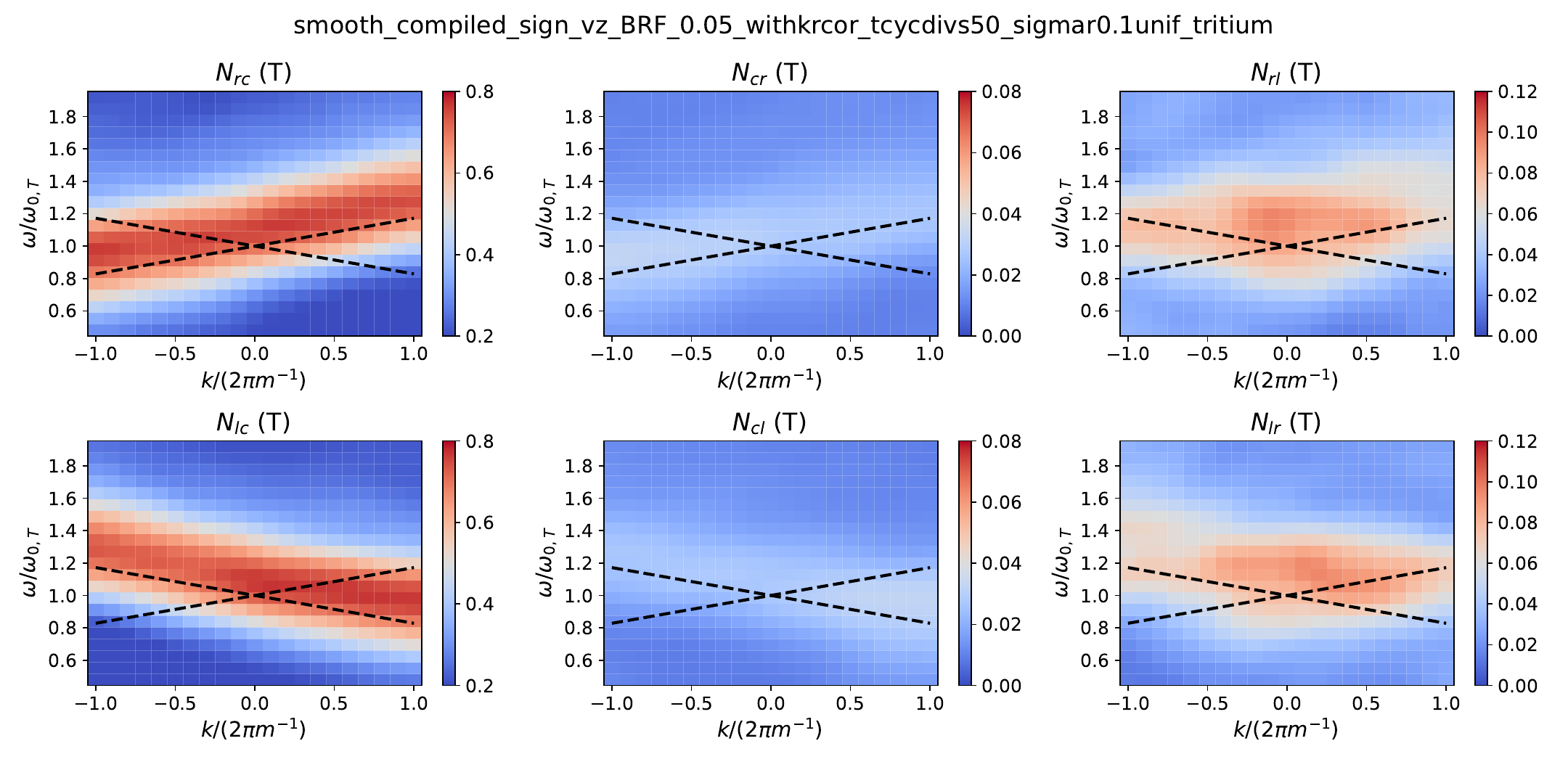}
    \caption{Smoothed and dimensionless RF rates, $\bar{N}_{ij}$, as a function of $k,\omega$ for tritium in TRMF (with induced electric field).
    The overlaid dashed black lines indicate the theoretical resonance condition for right- and left-going particles.}
     \label{fig: smoothed RF rates T, TRMF}
\end{figure}

\begin{figure}
    \centering
    \includegraphics[clip, trim=0.0cm 0.0cm 0.0cm 1.0cm, width=1.0\linewidth]{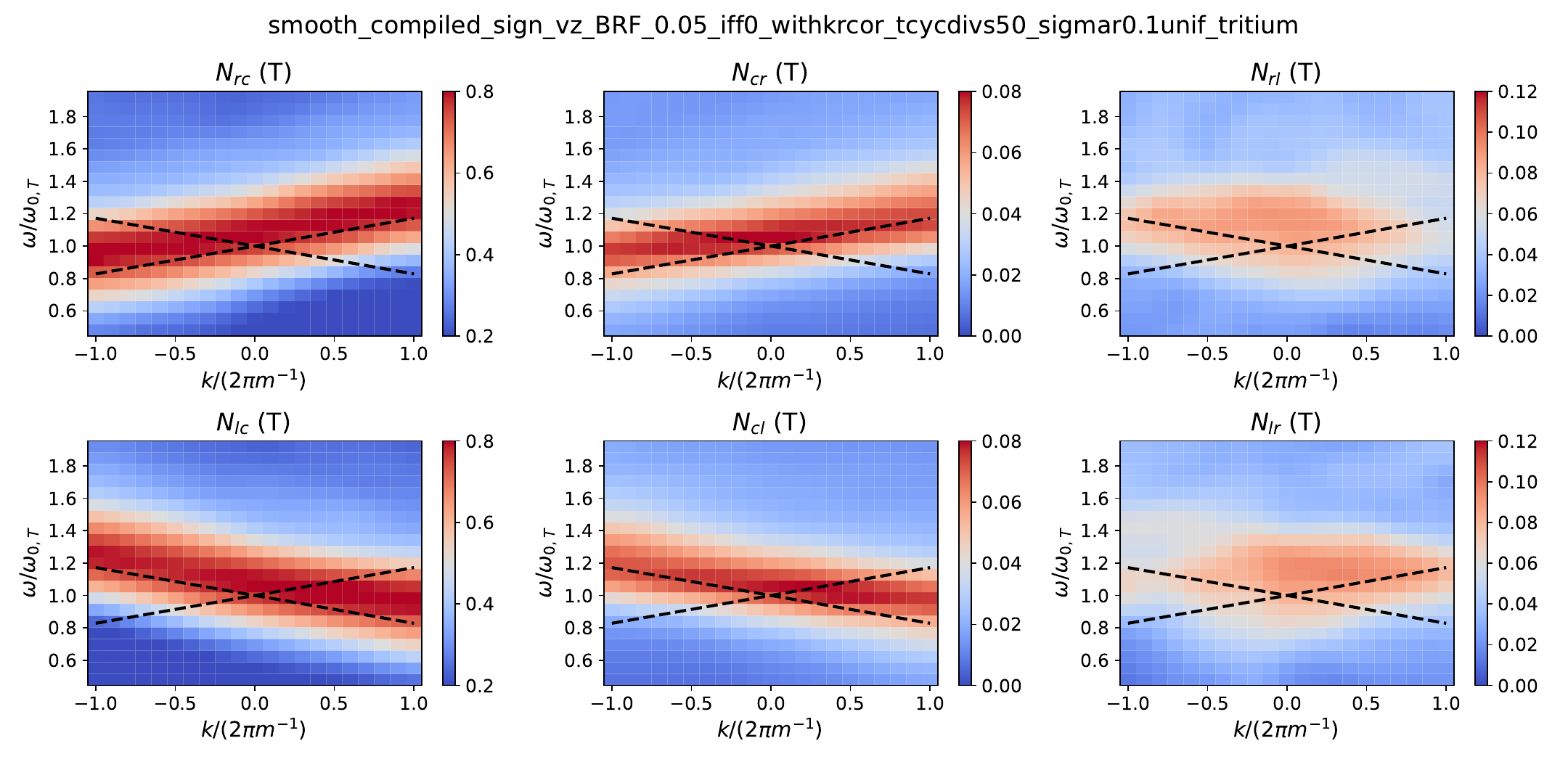}
    \caption{Smoothed and dimensionless RF rates, $\bar{N}_{ij}$, as a function of $k,\omega$ for tritium in TRMF--noE. 
    The overlaid dashed black lines indicate the theoretical resonance condition for right- and left-going particles.}
    \label{fig: smoothed RF rates T, TRMF iff=0}
\end{figure}

\begin{figure}
    \centering
    \includegraphics[clip, trim=0.0cm 0.0cm 0.0cm 1.0cm, width=1.0\linewidth]{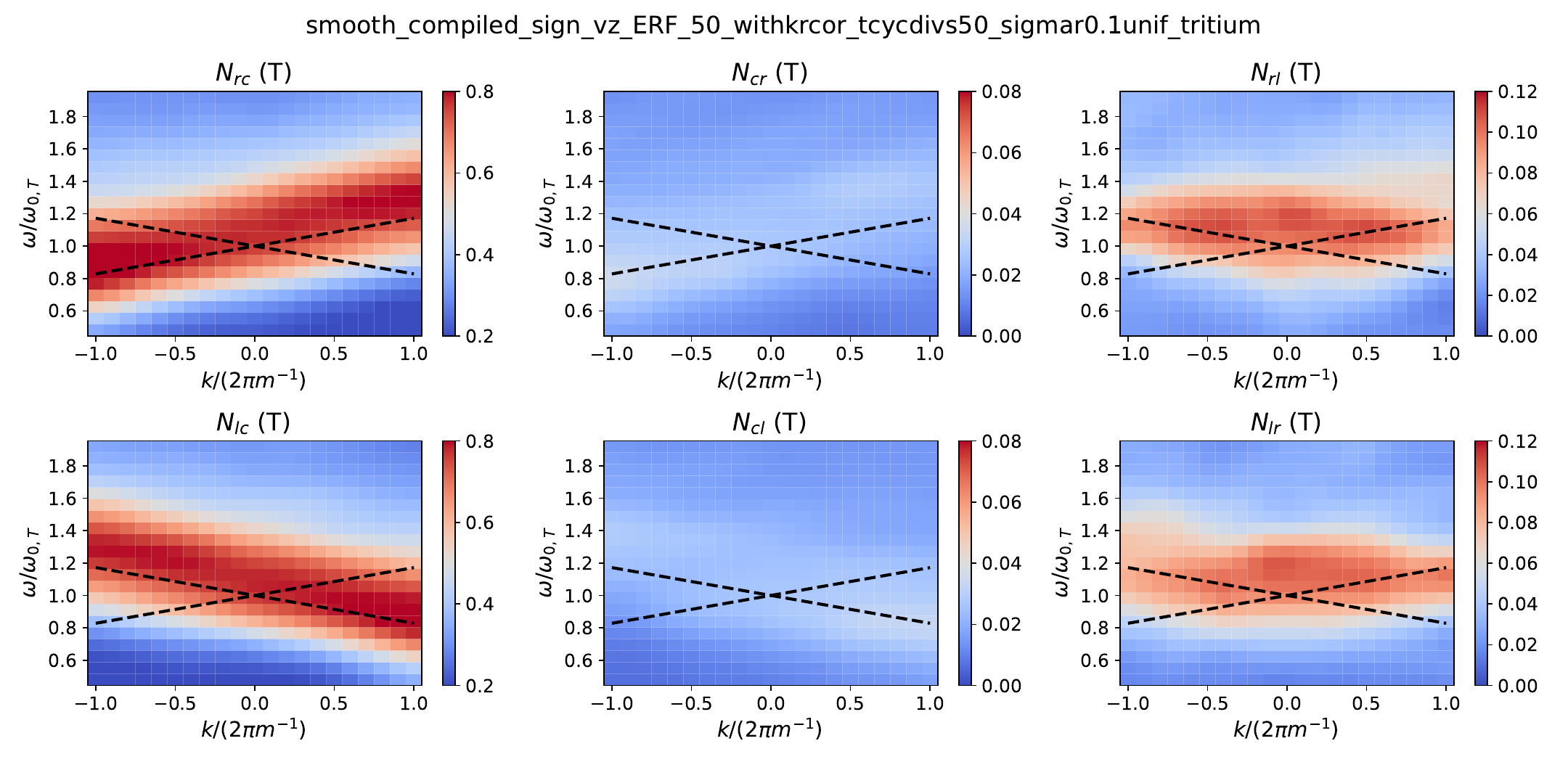}
    \caption{Smoothed and dimensionless RF rates, $\bar{N}_{ij}$, as a function of $k,\omega$ for tritium in TREF. 
    The overlaid dashed black lines indicate the theoretical resonance condition for right- and left-going particles.}
 \label{fig: smoothed RF rates T, TREF}
\end{figure}

\begin{figure}
    \centering
    \includegraphics[clip, trim=0.0cm 0.0cm 0.0cm 1.0cm, width=1.0\linewidth]{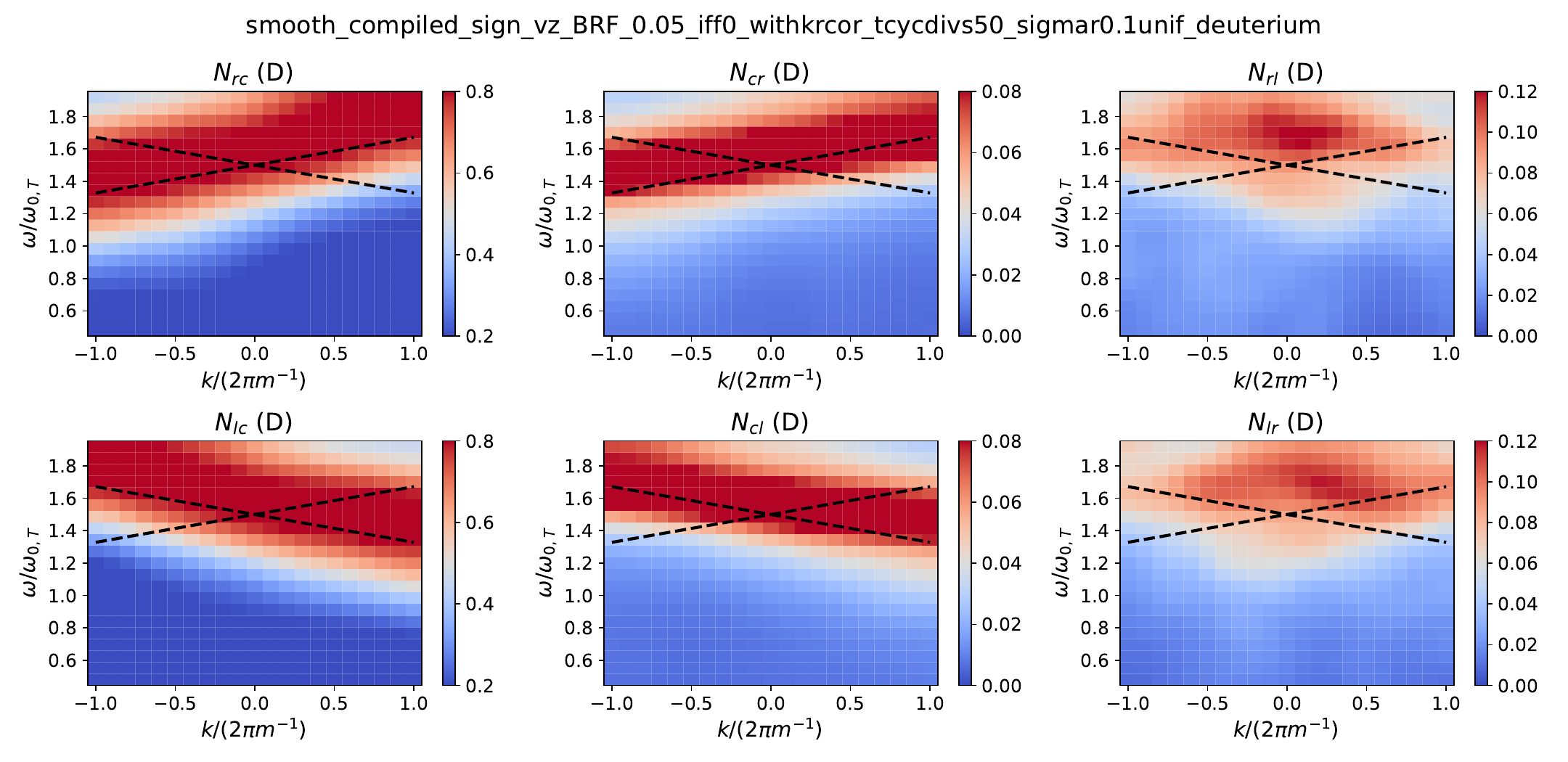}
    \caption{Smoothed and dimensionless RF rates, $\bar{N}_{ij}$, as a function of $k,\omega$ for deuterium in TRMF--noE. 
    The overlaid dashed black lines indicate the theoretical resonance condition for right- and left-going particles.}
 \label{fig: smoothed RF rates D, TRMF iff=0}
\end{figure}

The RF-induced rates for the four dominant transitions, $\bar{N}_{rc},\bar{N}_{cr}, \bar{N}_{lc}, \bar{N}_{cl}$, are found to be concentrated along straight lines in the ($k,\omega$) parameter space.
These lines are associated with the cyclotron resonance condition, which results from the compensation between the Doppler frequency detuning (in the moving frame of each population) and the RF wave phase velocity, as discussed in Sec.~\ref{sec: fields}.
The overlaid dashed black lines in the figures indicate the resonance condition, with $v_z$ replaced by the mean axial velocity within the loss cones
\begin{eqnarray}
\bar{v}_{z,LC}=\frac{\intop_{LC}\,f_{MB}\left(\mathbf{v}\right)\,v_{z} \,d^{3}v}{\intop_{LC}f_{MB}\,\left(\mathbf{v}\right)\,d^{3}v}. 
\end{eqnarray}
Here, $f_{MB}\left(\mathbf{v}\right)=\pi^{-3/2}\, v_{th}^{-3} \exp(-\mathbf{v}^2/v_{th}^2)$ is the Maxwell-Boltzmann distribution function, where the integral is taken only over the loss cone region of the velocity space.
One finds that
\begin{eqnarray}
    \bar{v}_{z,LC}=\pi^{-1/2} \left( 1 +\sqrt{1-R_m^{-1}} \right) v_{th}, 
\end{eqnarray}
where for the considered mirror ratio $R_m=5$ the solution is $\bar{v}_{z,LC}=1.07 \,v_{th}$.
Notably, the right and left loss cones have mean axial velocities in opposite directions, resulting in resonance lines with opposite slopes.

It can also be seen that the theoretical lines appear at slightly lower frequencies compared to the peak rate values. 
This can be explained by the increase in magnetic field experienced by ions as they move away from the mirror midplane, leading to higher cyclotron frequencies.
For deuterium, the transition-rate maps in Fig.~\ref{fig: smoothed RF rates D, TRMF iff=0} are shifted to higher frequencies relative to tritium, reflecting the inverse dependence of the cyclotron frequency on ion mass.
\section{Rate Equations Model}
\label{sec: rate eqs}

In previous work, we generalized the semi-kinetic rate equations model for the MM system [\cite{miller2021rate}], which included only particle transmission between neighboring cells through the loss cones and Coulomb scattering within each cell, and added additional terms that represent driven transport induced by external RF fields [\cite{miller2023rf}].
Here, we adapt the rate-equations model and further generalize it to include the transitions between left- and right-going populations, which were neglected in [\cite{miller2023rf}]. 
The resulting rate model, therefore, includes all six RF-induced transitions $\nu_{RF,ij}$ as defined in Eq.~(\ref{eq: RF rate}) and estimated from the single particle Monte Carlo simulations.

The generalized model then reads:
\begin{eqnarray}
\label{Eq: dn_c_dt} \dot n_{c}^i &=&  \nu_{s} \left[(1-2 \alpha) (n_{l}^i + n_{r}^i) - 2 \alpha n_{c}^i\right]  \\
                        &&  - \left( \nu_{RF,cl} + \nu_{RF,cr} \right) n_{c}^i + \nu_{RF,lc} n_{l}^i + \nu_{RF,rc} n_{r}^i \nonumber \\ 
\label{Eq: dn_tL_dt} \dot n_{l}^i &=&  \nu_{s}\left[\alpha (n_{r}^i+n_{c}^i) - (1-\alpha) n_{l}^i\right] - \nu_{t} n_{l}^i + \nu_{t} n_{l}^{i+1} \\ 
                        &&  -  \left( \nu_{RF,lc} + \nu_{RF,lr} \right) n_{l}^i + \nu_{RF,cl} n_{c}^i + \nu_{RF,rl} n_{r}^i \nonumber \\ 
\label{Eq: dn_tR_dt} \dot n_{r}^i &=& \nu_{s} \left[\alpha (n_{l}^i +n_{c}^i) - (1-\alpha) n_{r}^i\right]- \nu_{t} n_{r}^i +  \nu_{t} n_{r}^{i-1} \\
                        &&  -  \left( \nu_{RF,rc} + \nu_{RF,rl} \right) n_{r}^i + \nu_{RF,cr} n_{c}^i + \nu_{RF,lr} n_{l}^i  \nonumber, \; 
\end{eqnarray}
where $\nu_s$ is the ion-ion Coulomb scattering rate that roughly scales with density and temperature as $\propto n/T^{3/2}$;  $\nu_t = f_t v_{th}/l$ the inter-cell transmission rate, where $v_{th}$ is the thermal velocity of the ions and $l$ is the length of the mirror cell and the ambipolar factor is $f_t=\left( T_i + T_e  \right)/T_i=2$; and finally $\alpha$ is the normalized loss-cone solid angle.
For the mirror ratio considered $R_m=5$ the loss-cone angle is $\theta_{LC} =  \arcsin \left( R_{m}^{-1/2} \right) \approx 26.5 ^{\circ} $ giving the normalized loss-cone solid angle $\alpha = \sin^{2} \left(\theta_{LC}/2\right) \approx 0.053$; Therefore, the relative size of each loss-cone is $\alpha$ and the confined section $1-2\alpha$.
Among the different thermodynamic scenarios studied in [\cite{miller2021rate}], we considered here the simple isothermal scenario, in which $\nu_s,\nu_t$ are constant across all cells, since the external RF fields dominate the scattering rather than collisions.

We close the rate equations model by imposing the following boundary conditions: a constant total density in the throat of the central fusion cell (the leftmost cell),
\begin{equation}
    n_c^1+n_{l}^1+n_{r}^1 =n_0 =\mathrm{const}, 
\end{equation}
where $n_0$ is the density in the central cell, and a free-flow boundary condition at the exit of the outermost cell,
\begin{equation}    
    \nu_t n_l^{N+1}=0,
\end{equation}
corresponding to the absence of left-going flux from outside the system.

We solve the rate equations (\ref{Eq: dn_c_dt}-\ref{Eq: dn_tR_dt}) for the steady-state solution, \ie $\dot{n}=0$ for all cell populations in all cells.
The outgoing flux between two neighboring cells (\eg $i\rightarrow i+1$) is proportional to $\phi_{i,i+1} \propto v_{th} \left( n_{r}^i - n_{l}^{i+1} \right)$. 
In a steady state, by definition, all inter-cell fluxes are the same and denoted by $\phi_{ss}$. 
The system confinement time scales inversely with the steady-state flux,  $\tau \propto 1/\phi_{ss}$, which we aim to maximize to satisfy the Lawson criterion [\cite{lawson1957some, wurzel2022progress}].
It is noted that in the absence of RF fields, the confinement time scales linearly with the number of mirror cells. 
However, even under optimized and optimistic conditions, fusion-relevant confinement would require an impractically large number of cells [\cite{logan1972multiple, makhijani1974plasma, miller2021rate}].

\begin{figure}
    \centering
    \includegraphics[clip, trim=0.0cm 0.0cm 0.0cm 1.0cm, width=0.8\linewidth]{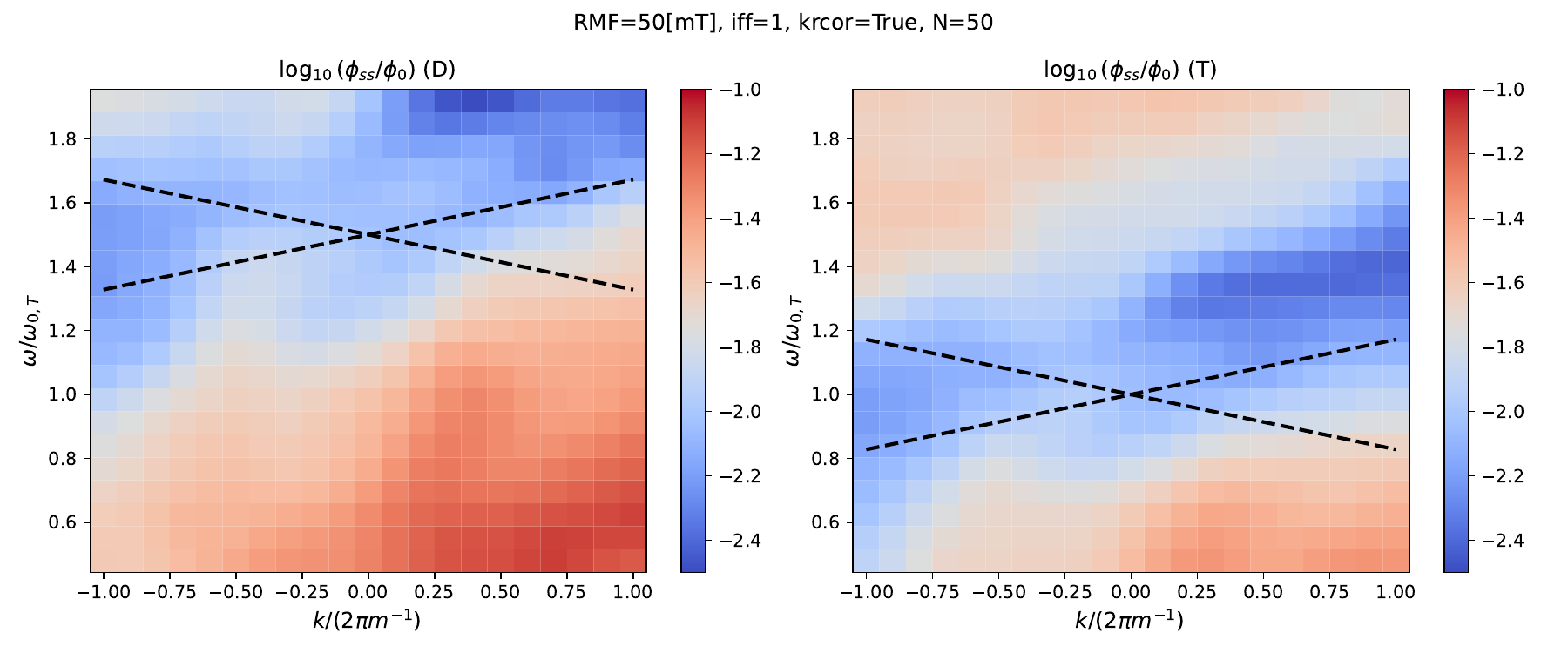}
    \includegraphics[clip, trim=0.0cm 0.0cm 0.0cm 1.0cm, width=0.8\linewidth]{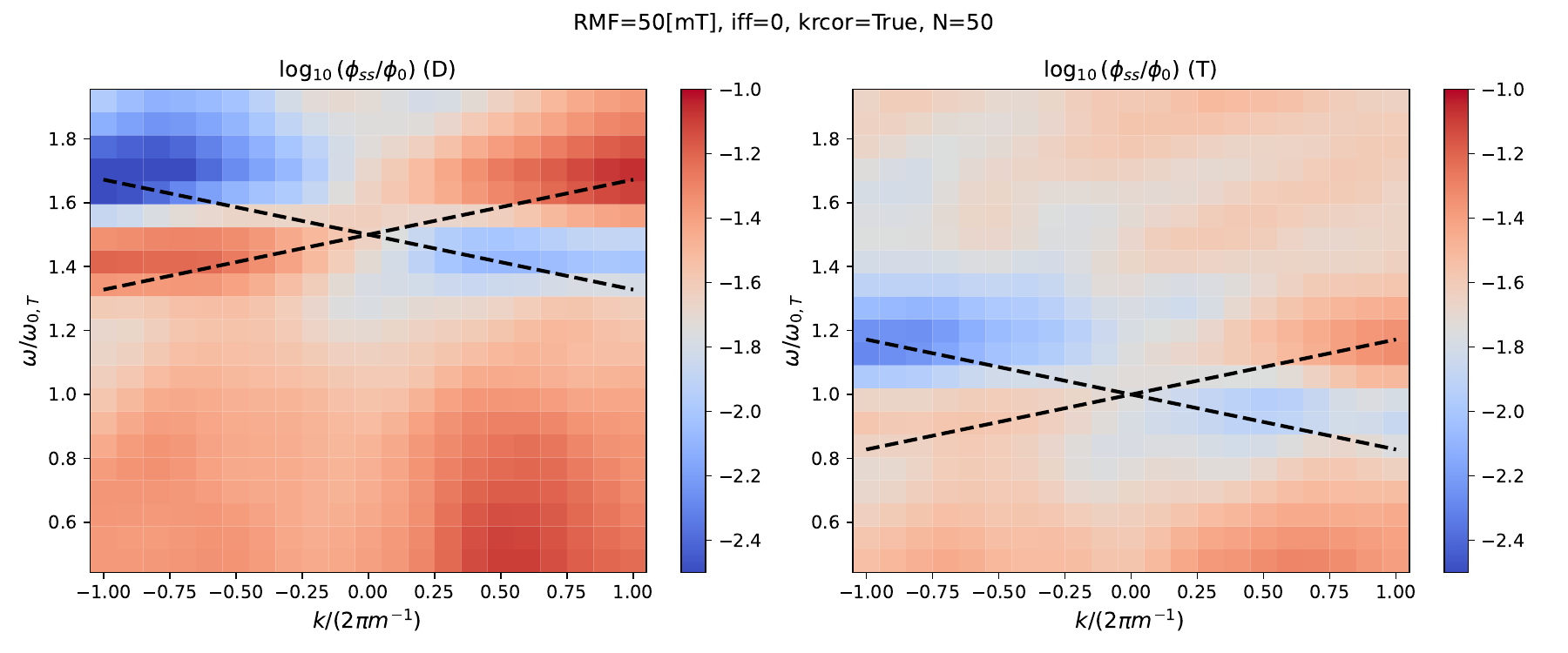}
    \includegraphics[clip, trim=0.0cm 0.0cm 0.0cm 1.0cm, width=0.8\linewidth]{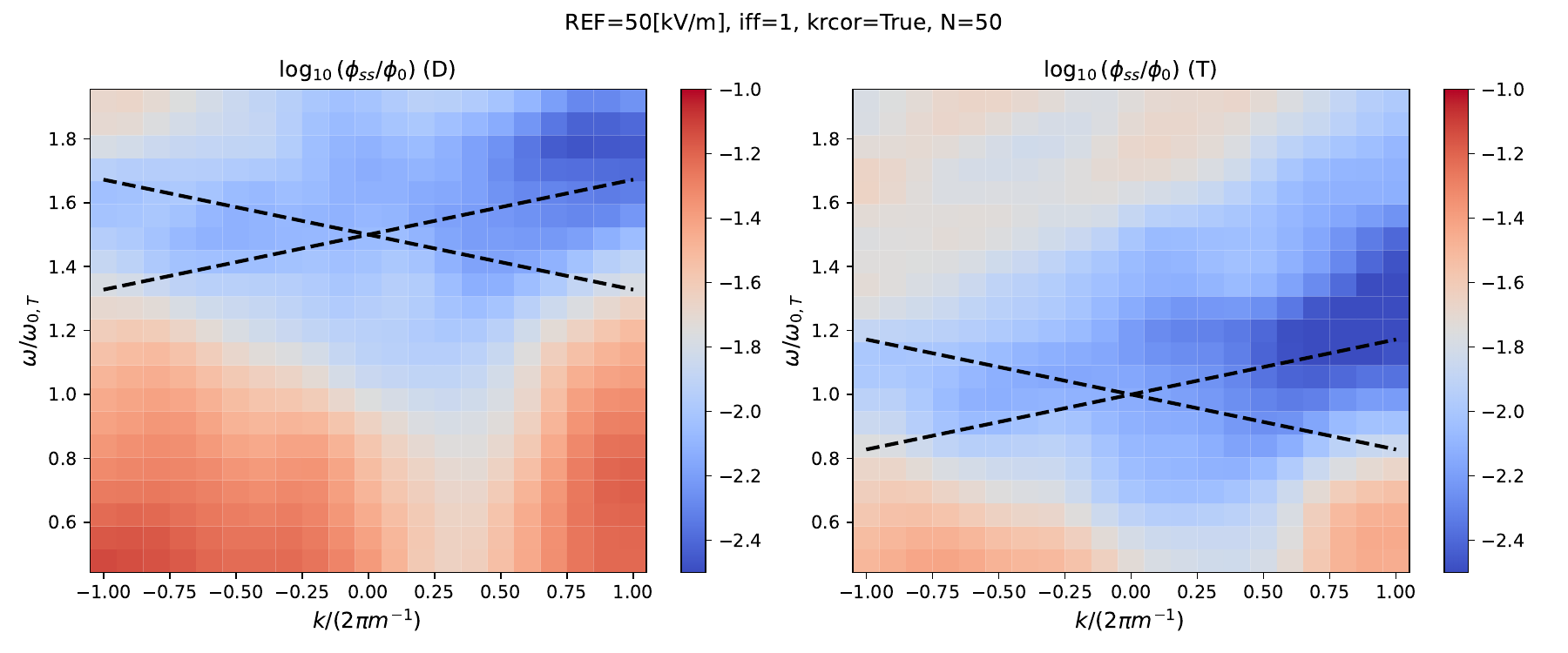}
    \caption{Steady state flux in an MM system with $N=50$ cells as a function of $k$ and $\omega$ for deuterium (left) and tritium (right). 
    The RF schemes are TRMF (top), TRMF--noE (center) and TREF (bottom). 
    The overlaid dashed black lines indicate the theoretical resonance condition for right- and left-going particles.}
     \label{fig: steady state flux of k}
\end{figure}

It is noted that because our rate-equations model describes a single ion species, it cannot directly capture the dynamics of a deuterium–tritium plasma.
However, in regimes where Coulomb scattering rates are negligible compared with the RF-induced transition rates, the two ion species may be treated as non-interacting.
Under this assumption, the rate equations are solved separately for deuterium and tritium, using RF transition rates obtained independently for each species from the single-particle simulations presented in Sec.~\ref{sec: single particle}.
Figure~\ref{fig: steady state flux of k} shows the results for deuterium and tritium under different RF schemes and for various values of $k$ and $\omega$ at a fixed number of cells, $N=50$. 
The theoretical resonance lines, as defined in Sec.~\ref{sec: single particle}, are indicated by dashed lines.
The fluxes shown in the figures are normalized by the single-mirror flux, $\phi_0 = n_0 v_{th}$, which provides a convenient reference for illustrating the flux reduction achieved by the various plugging schemes considered in the simulations.
This normalization differs from that used in [\cite{miller2021rate, miller2023rf}], where the flux was normalized by the maximum value compatible with satisfying the Lawson criterion [\cite{lawson1957some, wurzel2022progress}].
Here, however, we use a simpler normalization to emphasize the primary objective of flux reduction relative to the single-mirror value.

From the figure, it can be seen that the confinement improvement, quantified by the flux reduction, in all three RF scenarios over the parameter range explored in the simulations lies between one and approximately $2.5$ orders of magnitude.
Along the resonant line corresponding to the right-going (outgoing) particles, both the TRMF and TREF configurations exhibit the strongest confinement enhancement, approaching a reduction of about $2.5$ orders of magnitude.
In contrast, for the TRMF--noE configuration, the optimal parameter region lies along the opposite resonant line, corresponding to resonance with right-going (incoming) particles.
This non-intuitive result is discussed in detail in the following section.

\begin{figure}
    \centering
    \begin{minipage}{0.49\textwidth}
        \centering
        \includegraphics[clip, trim=0.0cm 0.0cm 0.0cm 0.8cm, width=\linewidth]{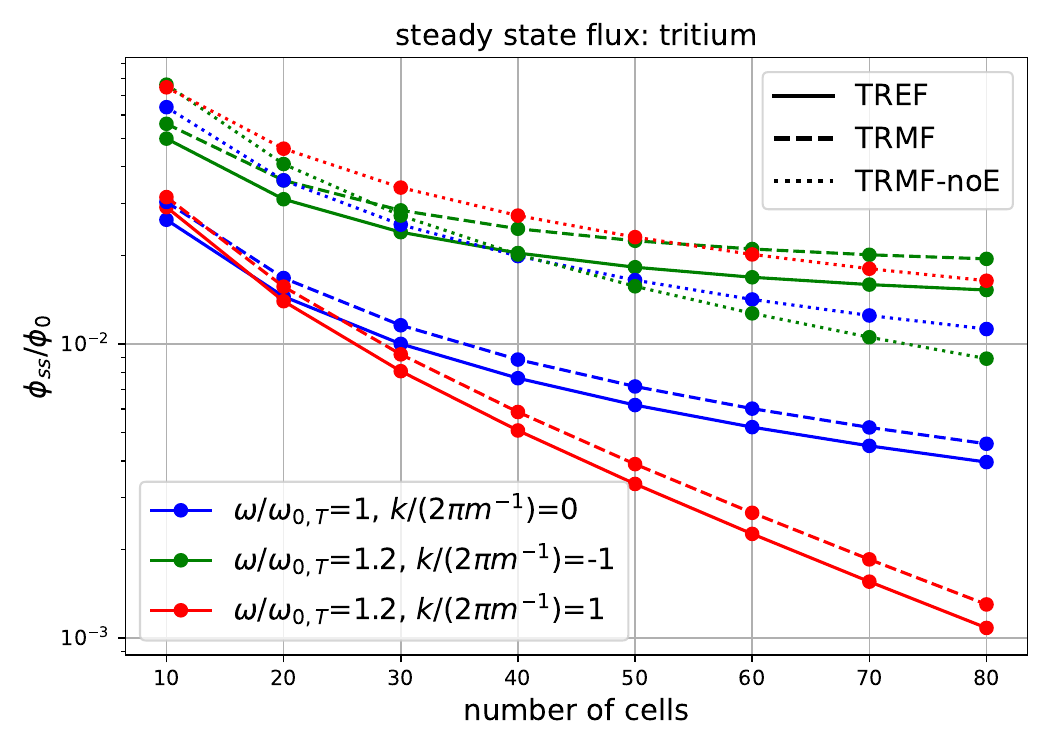}
    \end{minipage}
    \caption{Steady state flux for tritium as a function of the number of MM cells for three different parameter sets (see legend).
    Line styles denote TREF (solid), TRMF (dashed), and TRMF--noE (dotted).}
    \label{fig: steady state flux of N}
\end{figure}

In addition, outside the resonant lines, both the TRMF and TREF schemes provide stronger confinement enhancement than the TRMF--noE configuration, resulting in greater suppression of the steady-state flux, as discussed in the following section.
As a rough estimate, the axial flux must be reduced by at least four orders of magnitude relative to $\phi_0$ for an MM system to become viable as a fusion device [\cite{miller2023rf}].
Consequently, while the present technique provides substantial confinement enhancement, additional plugging mechanisms will be required for fusion-relevant applications, or alternatively, operation in a different parameter regime or the adoption of a modified system design.

Fig.~\ref{fig: steady state flux of N} examines the scaling of the steady-state flux with the number of MM cells, $N$, for selected RF configurations, as indicated in the legend.
In the resonant cases, the flux decays approximately exponentially with $N$.
However, the decay rate associated with resonance of right-going (outgoing) particles, corresponding to the TREF and TRMF schemes (red curves), is significantly larger than that associated with resonance of left-going particles in the TRMF--noE configuration (green curve).
This indicates that, even under optimal resonant conditions for each scheme, the TRMF--noE configuration is less effective at suppressing the axial flux than the other two RF schemes.
Nonetheless, despite its weaker confinement performance, the TRMF--noE configuration offers a critical practical advantage, \ie dramatically lower power consumption, as discussed in the next section.
\section{Discussion}
\label{sec: discussion}

\subsection{Power considerations}

When the RF field has a non-zero electric component, energy is deposited into the plasma.
Estimating this energy cost is therefore essential for assessing the feasibility of each technique and for guiding the choice of an appropriate tool for a given system, problem, and parameter regime.
In the single-particle simulations, the RF power can be estimated by tracking the average change in particle energy between the initial energy, $E_i$, and the final energy, $E_f$, over the simulation time $\tau_{th}$, as described in  Sec.~\ref{sec: single particle}.
The averaging is performed separately for particles originating from each of the three populations.
The average power associated with each population is then defined as
\begin{eqnarray}
    P_j =\frac{ \left<  E_f - E_i  \right>_j}{\tau_{th}}
\end{eqnarray}
where $j \in \{ c,r,l\}$, corresponds to confined, right-going, and left-going particles.

\begin{figure}
    \centering
    \begin{minipage}{0.49\textwidth}
        \centering
        \includegraphics[clip, trim=15.0cm 0.0cm 0.0cm 0.8cm, width=\linewidth]{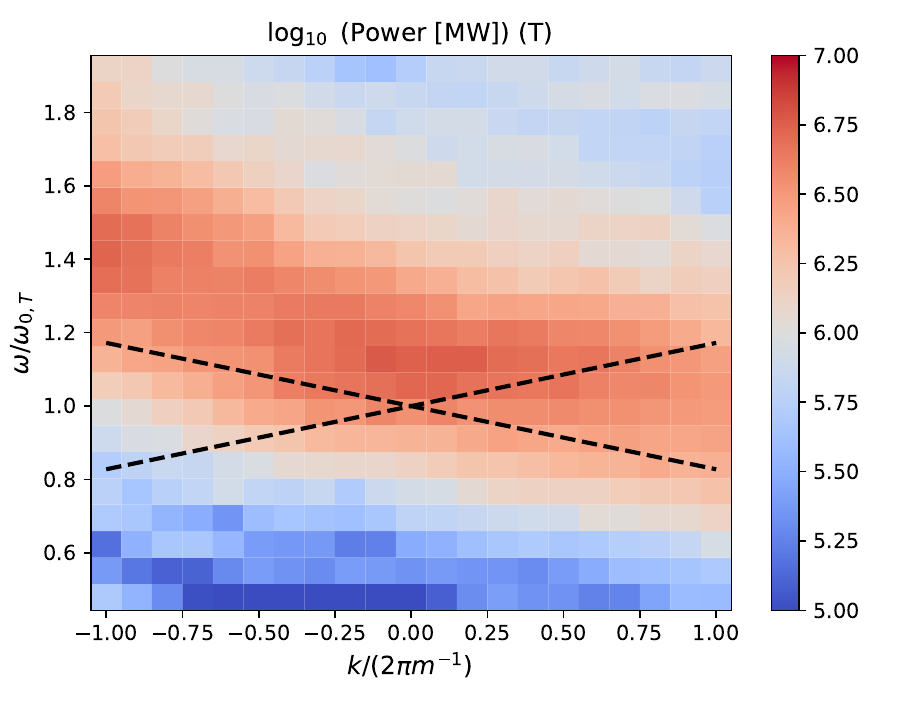}
    \end{minipage}
    \hfill
    \begin{minipage}{0.49\textwidth}
        \centering
        \includegraphics[clip, trim=15.0cm 0.0cm 0.0cm 0.8cm, width=\linewidth]{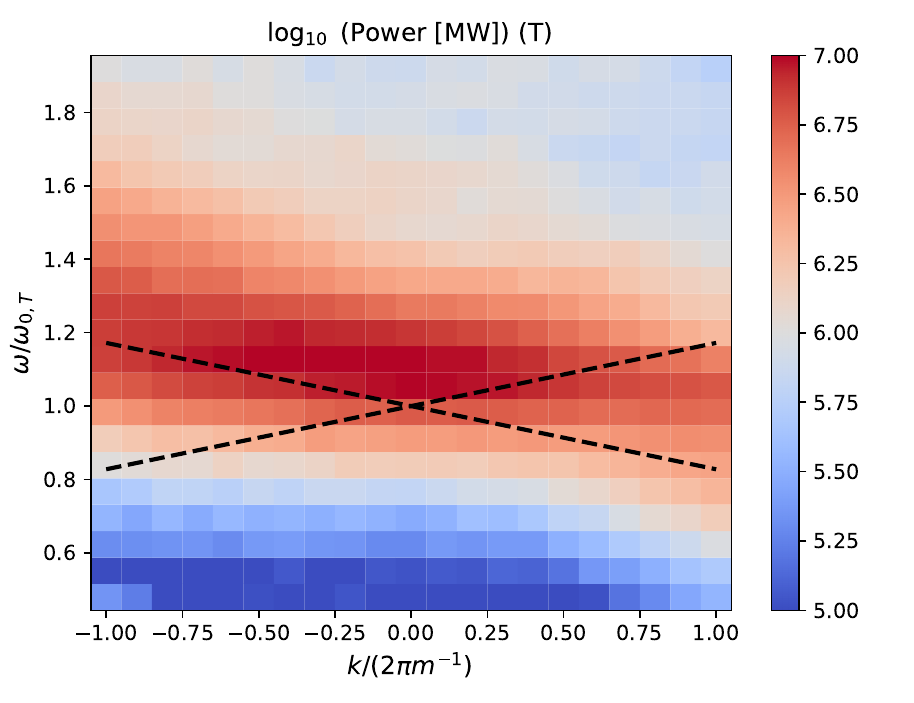}
    \end{minipage}
    \caption{RF power estimate in the MM for tritium as a function of $k,\omega$, for TRMF (left) and TREF (right). For $N=50$ cells.}
 \label{fig: RF power estimates}
\end{figure}

We then use each steady-state solution of the rate-equation model obtained in Sec.~\ref{sec: rate eqs} to estimate the total RF power deposited in the MM system as
\begin{eqnarray}
    P&=&V_{cell}\sum_{i=1}^{N}\sum_{j=\{ c,r,l\}}n_{j}^{i}P_{j}
\end{eqnarray}
where $V_{cell}=\pi R^2l$ is the plasma volume of a single cell, and $n^i_j$ is the steady-state density profile for  population $j$ in MM cell $i$ obtained from the solution to the rate equations.
Figure~\ref{fig: RF power estimates}  presents the power cost estimates associated with the steady state solutions in Fig.~\ref{fig: steady state flux of k} for tritium for both  TRMF (with induced electric field) and with TREF.
It is shown that the resulting power estimates exceed $10^5~\mathrm{MW}$ in all cases and are therefore unrealistically large, surpassing by many orders of magnitude the typical input or output powers of a fusion device. 

The enormous energy cost associated with the TREF and TRMF schemes may, at least in part, be attributed to the deposition of RF energy into particles that are already magnetically confined. 
Although the intent is to selectively target particles in a single (outgoing) loss cone, the RF interaction instead couples to all particles with specific axial velocities. 
Such velocities are also present in the confined population, since trapped particles travel back and forth within each mirror cell, spending approximately half of their bounce period traveling outward, with axial velocities comparable to those of escaping particles. 
As a result, the traveling RF fields also interact with particles that are already confined. 
Since, in steady state, the confined population constitutes the largest fraction of the distribution due to its large solid angle, it dominates the power absorption. 
As a result, much of the RF power is dissipated as unnecessary heating, thereby reducing the overall energetic efficiency of these schemes.
This analysis suggests that RF methods that transfer perpendicular energy to ions may be unsuitable for dense plasmas, such as those found in magnetic mirror (MM) systems, at least in regions near the central (fusion) cell. 
However, for laboratory-scale systems with dilute plasmas or for the outer MM cells in fusion-scale systems, these methods may remain favorable due to their high efficiency, despite the associated energy cost.

It is important to note that these estimates represent an upper bound on the power consumption. 
In practice, delivering such an amount of energy from the RF coils to the plasma would be unrealistic. 
For a given RF power, the coils are therefore expected to interact effectively with only a fraction of the particle population, thereby reducing the overall efficiency. 
This interaction may result either in a modest increase in perpendicular energy distributed across the entire population or in a substantial gain in the perpendicular energy for a limited subset of particles that become trapped, while the remaining particles are largely unaffected. 
In particular, collective effects in a dense plasma may screen the electric fields near the mirror axis or modify their structure, such that the net rate of energy gain is substantially lower than that predicted by the vacuum-field, single-particle model employed here.
Consequently, the field amplitudes or configurations assumed in the simulations likely overestimate the effective power transfer under realistic dense plasma conditions.

The other limiting scenario of TRMF arises when the plasma completely screens the induced electric fields, effectively behaving as a perfect dielectric medium. 
We refer to this limit as TRMF--noE. 
In light of the preceding power considerations, this scenario is of particular interest because the absence of penetrating electric fields implies that no significant RF power is deposited in the plasma beyond that required to maintain field screening.
Reality could lie somewhere in between these cases, so the solution of the rate equations, although not fully self-consistent, can be viewed as a sensitivity test of the RF fields configuration as experienced by the ions deep inside the plasma.
Detailed simulations, including collective and kinetic effects such as MHD, Vlasov, or (hybrid) PIC, are needed to better understand the mutual interaction between the external RF fields and the plasma particles and to realistically estimate the measure of the electric field penetration into the plasma and the consequent deposited power and plugging efficiency.
Such simulations are beyond the scope of the current paper.

Similarly, a key open question concerns the penetration of the magnetic field into the plasma.
Classical RMF theory predicts a threshold condition for magnetic field penetration into the plasma [\cite{hugrass1981numerical, milroy1999numerical}].
Applying this framework to our example suggests that the TRMF amplitudes considered above satisfy the threshold criterion, so the magnetic field is expected to overcome skin-depth screening and penetrate the plasma core.
However, the assumptions underlying this theory are not satisfied in the present regime.
In particular, the theory is derived for cold plasmas and higher-frequency operation, effectively assuming immobile ions, whereas our system considers a hot plasma and frequencies near the ion cyclotron frequency.
Thus, the classical threshold theory is not formally applicable in this case, and advanced simulations or experiments, which are beyond the scope of this work, are required for a rigorous assessment of magnetic-field penetration.
Nevertheless, the feasibility of RMF penetration is well established, particularly in RMF-driven FRCs [\cite{jones1999review, furukawa2019verification, polzin2020state, cohen2023laboratory}], providing indirect evidence that a suitable parameter regime may exist for TRMF plugging in MM systems.

Another caveat concerns the plasma temperature. 
In solving for the steady state of the rate equations in Sec.~\ref{sec: rate eqs}, the density and temperature profiles are, by definition, assumed to remain constant in time. 
However, the external RF fields deposit a finite power that may exceed the plasma’s radiative cooling rate, thereby violating this steady-state assumption. 
A more consistent model, which lies beyond the scope of the present analysis, would need to account for both effects. 
Consequently, within the present framework, TRMF--noE appears more self-consistent than the other RF schemes, as it transfers no significant net energy to the plasma.
Next, we discuss further aspects of this scenario.

\subsection{TRMF--noE: Trapping mechanism}
\label{sec: TRMF-noE}

In the TRMF case that includes the induced electric field, both the plugging performance and the energetic cost are similar to those obtained for TREF, which we have associated with the injection of perpendicular energy that removes particles from the loss cones. 
Thus, although the TRMF dynamics is more complex as the induced electric field is directed along the axial direction, its trapping mechanism may be related to the acquisition of transverse energy in a qualitatively similar manner to TREF. 
In contrast, in the limiting TRMF--noE scenario, no net energy is injected into the plasma, since in the absence of an electric field, no work is done on the particles. 
This raises the question of what physical mechanism is responsible for particle plugging in this case.

To answer this question, we first look at a simple RMF (see Eq.~\ref{B_RMF}) in a perfect stationary ion-cyclotron resonance, \ie $k=0$ and $\omega=\omega_0$, where $\omega_0=qB_0/m$, and note that in coordinates rotating with the particle, the magnetic field always points radially at a fixed direction, which we define as $\hat{x}'$.
Here, we neglected the relativistic effects because the particle (cyclotron) velocity is small compared to the speed of light.
Now, assume the particle moves in the axial $z-$direction, which does not change in the rotating frame, \ie $\hat{z}'=\hat{z}$, the Lorentz force in the particle's rest frame, points to the $\hat{y}'$ direction.
The cyclotron motion in the moving frame is nothing but a rotation around $\hat{x}'$ at the frequency $\omega_1=qB_\text{RF}/m$. 
In other words, the particle rotates in the $\hat{x}'-\hat{z}'$ plane, periodically switching its total (conserved) energy between 
the $\hat{x}'$ and $\hat{z}'$ directions, \ie between axial and perpendicular energies.
When the resonance is not perfect, and the particle velocities are distributed thermally, the overall motion involves stochastic switching between transverse and axial motions.
This effect can be interpreted as an effective elastic collision.

A direct picture of the dynamics is provided by an analytical solution of the charged-particle equation of motion in a constant static magnetic field with a simple, perfectly resonant RMF.
As before, we consider a uniform magnetic field with an additional rotating component in a perfect cyclotron resonance of the unperturbed magnetic field
\begin{equation}
    \mathbf{B}=B_{RF}e^{-i\omega_0 t}\left(\hat{x}-i\hat{y}\right)+B_{0}\hat{z},
\end{equation}
and neglect the effect of the induced electric field.
For a particle with initial velocity of $\mathbf{v}_{0}=\left(v_{\perp,0},0,v_{z,0}\right)$, one finds that the solution for the particle velocity under the influence of the Lorentz force, $m\dot{\mathbf{v}}=q\mathbf{v}\times\mathbf{B}$, reads
\begin{align}
    v_{x}+iv_{y}	&= \left(v_{\perp,0}+iv_{z,0}\sin\left(\omega_{1} t\right)\right)e^{-i\omega_{0}t} \label{eq: vx_ivy} \\
    v_{z}	&= v_{z,0}\cos\left(\omega_{1} t\right).
\end{align}

By multiplying Eq.~(\ref{eq: vx_ivy}) by $e^{i\omega_{0}t}$, the dynamics of the particle in the RMF rotating frame, can be interpreted as an effective cyclotron rotation in the slow frequency, $\omega_1$, in a plane determined by the $\hat{z}$-direction and the rotating radial vector.
The particle energy, and thus, the amplitude of the velocity, $\left|\mathbf{v}\left(t\right)\right|$, remains constant because in the absence of an electric field, the motion is governed solely by magnetic fields, which perform no work.
Extending this picture to a particle in a magnetic mirror under the influence of an RMF, the effective cyclotron motion in the rotating plane, as demonstrated above for a uniform magnetic field, induces a periodic motion in phase space that can cross the loss cone, \ie transitions between the escaping and confined populations.
This picture can be considered as an RF-induced, effective collision mechanism with a characteristic switching frequency $\omega_1$.

Now, let us reexamine the rate equations model.
As shown in Fig.~\ref{fig: steady state flux of k}, TRMF and TREF achieve their strongest confinement along the resonant line for right-going (outgoing) particles, whereas TRMF--noE is optimized along the opposite resonant line, \ie for left-going (incoming) particles.
This indicates that the plugging mechanism in TRMF--noE differs from those in both TRMF and TREF.
It seems that this difference is related to perpendicular energy deposition that happens in TRMF and TREF but not in TRMF--noE, due to energy conservation in the latter.
For TREF, it is a direct conclusion from the single-particle simulations (see Fig. 2 in [\cite{miller2023rf}]).
The trapping mechanism of TRMF is more complex, but by comparing the RF rates in the schemes TRMF (Fig.~\ref{fig: smoothed RF rates T, TRMF}) and TRMF--noE (Fig.~\ref{fig: smoothed RF rates T, TRMF iff=0}), it can be seen that the values of $\bar{N}_{rc}$ are similar, while $\bar{N}_{cr}$ is much smaller in TRMF. 
One concludes that the induced electric field in the TRMF scenario suppresses the de-trapping effect of the rotating magnetic field by altering the energy of trapped particles, thereby increasing the ratio of the perpendicular to the longitudinal energy.

Interpreting the numerical results together with the analytical solution for the simple RMF trajectory, we suggest that the plugging mechanism in the TRMF--noE scenario can be understood as an effective diffusion that is induced by the rotating magnetic field.
The interpretation of an effective diffusion is also supported by the structure of the rate equations.
In a hypothetical symmetric case in which all RF transition rates are equal after normalizing by the solid angle of the target population, \eg $\nu_{RF,cr}=\alpha\nu_{RF}$ and $\nu_{RF,rc}=(1-2\alpha)\nu_{RF}$ for some constant $\nu_{RF}$, the RF terms in Eqs.~(\ref{Eq: dn_c_dt}-\ref{Eq: dn_tR_dt}) play a role equivalent to that of Coulomb scattering.
In this limit, $\nu_{RF}$ can be added to the scattering rate, $\nu_s$, without altering the form of the equations.
More generally, in a partially symmetric case where transitions between any two populations occur at comparable rates, the dynamics can be viewed as governed by an effective diffusion (or mixing) between populations. 
This situation is analogous to the state of detailed balance in thermodynamics.
Therefore, a possible outcome of the RMF--noE scenario (with $k=0$) is a strong mixing between phase-space populations in the MM sections combined with low collisionality (\ie high temperature and low density) in the central cell, which is favorable for sustained fusion-relevant plasmas.
In other words, in the RMF--noE scenario, the RF-induced transition rates are controlled by the externally applied fields rather than by the plasma parameters.

Furthermore, when the RF trapping rates are asymmetric, we can achieve right-left selectivity, making trapping more favorable for escaping particles than for returning particles.
Exploiting this effect, the plugging of the MM section can be based on the RF effect rather than on collisions.
This also decouples the confinement properties of the MM sections from collisionality constraints, allowing the plasma parameters (\ie temperature and density) in the central cell to be optimized for fusion performance. 
Such a scenario may open an avenue toward energetically affordable RF-based plugging in fusion-scale MM systems. 
For example, plasma parameters typically considered for D-T fusion, $n=10^{21}\,\mathrm{m^{-3}}$ and $k_{B} T=10 \mathrm{keV}$, the corresponding MFP is of the order of kilometers.
This regime is impractical for classical MM configurations, which operate most effectively when the Coulomb-scattering MFP, $\lambda=v_{th} / \nu_s$, is comparable to the mirror-cell length, $l$.
In principle, the mean free path can be reduced by increasing the plasma density and lowering the temperature, while keeping the magnetic pressure fixed.
However, this requires substantially longer confinement times to satisfy the Lawson criterion [\cite{lawson1957some,wurzel2022progress,miller2021rate}], while radial diffusion may represent a significant limitation in this regime.
Instead, methods that induce effective scattering and thereby reduce the overall effective mean free path, such as TRMF--noE, offer a promising alternative.
\section{Conclusions}
\label{sec: conclusions}

In this work, we propose and analyze a novel method for enhancing plasma confinement in multi-mirror systems using a traveling rotating magnetic field (TRMF).
Given the uncertainty in the extent of electric field penetration into the plasma, we consider two limiting scenarios, with and without induced electric fields, \ie TRMF and TRMF--noE, respectively.
Both scenarios are compared with the previously studied traveling rotating electric field (TREF) scheme.
We employ single-particle Monte Carlo simulations to quantify the RF-induced transition rates between the confined, right-going (escaping), and left-going (incoming) particle populations in a single MM cell. 
The simulation results are incorporated into a semi-kinetic model to calibrate the rate coefficients in the generalized rate equation model.
The rate equations are then solved to evaluate the steady-state particle fluxes across multiple cells over a range of the RF-parameters, $k$ and $\omega$.
We repeat these calculations for the three regimes, TRMF, TRMF--noE, and TREF, and estimated the RF power deposition from the simulated energy changes for each scenario.

The main result is that both TRMF and TRMF--noE, similar to TREF, can significantly reduce the outgoing particle flux, thereby enhancing the confinement time.
However, unlike TREF and TRMF, the recapture mechanism in TRMF--noE, analogous to an elastic collision, is based solely on phase-space mixing rather than on the direct injection of perpendicular energy.
Simulations show that the most effective regime corresponds to the mixing between incoming and trapped populations within each MM cell.
This regime is achieved through the interplay between a finite RF wavenumber and the Doppler shift that maintains the cyclotron resonance primarily for left-going particles.
In contrast, the transition rates of TREF and TRMF (with an induced electric field) peak along cyclotron resonance lines with Doppler shifts, enabling directional selectivity that preferentially traps right-going (escaping) particles. 
For optimized RF parameters in each scenario, we find the steady-state flux to be suppressed by several orders of magnitude relative to a single mirror and to depend exponentially on the number of MM cells. 
Deuterium and tritium exhibit similar behaviors, with resonance frequencies scaled by their masses.

For RF schemes with non-zero electric-field components, \ie TREF and TRMF, the power deposition in the plasma (\ie directional heating) is significant and therefore more relevant for dilute plasmas than for full-scale fusion machines. 
On the other hand, even though the TRMF--noE scheme is less effective at increasing axial confinement compared to the other schemes, it could be more physically relevant for fusion systems with dense plasma since it is energetically inexpensive.
Notably, even with non-optimal RF parameters, TRMF--noE can effectively reduce the MFP in the MM section as the phase space mixing mechanism analogously replaces the Coulomb scattering while the collisional MFP remains large enough, as required for MM systems.
It is also important to note that configurations with net plasma heating are inconsistent with steady-state solutions since the temperature should increase with the RF energy deposition, whereas the TRMF--noE scheme is self-consistent by construction.

Several important issues are left for future work. 
Most importantly, the present study neglects collective plasma effects and self-consistent RF–plasma interactions.
In realistic dense plasmas, however, the penetration and modification of the RF fields, as well as possible paramagnetic amplification or screening, must be addressed through extensive simulations that account for collective and kinetic effects. 
Finally, experimental validation is essential to determine whether the predicted RF-induced transition rates and plugging efficiencies persist under realistic plasma conditions and antenna geometries.

\section*{Data Availability Statement}
The data that support the findings of this study are available from the corresponding author upon reasonable request.

\section*{Funding}
This work was supported by the PAZI Foundation, Grant No. 2020-191; and the Israeli Institute for Fusion Research, the Israeli Ministry of Energy, Grant No. R-25-4. 
\section*{Declaration of Interests}
The authors report no conflict of interest.

\bibliography{references}

\section{Appendix: Leading order analysis of the traveling rotating fields}

In this appendix, we examine the validity and consistency of the approximation underlying Eqs.~(\ref{eq:TRMF_B})-(\ref{eq:TRMF_E}) for the TRMF scenario.
To this end, we substitute the proposed fields into Maxwell's equations and track the leading-order terms in $\left(\omega r/c\right)$ and $\left(kr\right)$.
The proposed electric and magnetic fields forms (as defined in Eqs.~(\ref{eq:TRMF_B})-(\ref{eq:TRMF_E})) are
\begin{eqnarray}
    \mathbf{B}	&=& B_{1}e^{i\varphi}\left[\left(\hat{x}-i\hat{y}\right)+k\left(y+ix\right)\hat{z}\right] \\
    \mathbf{E}	&=& B_{1}\omega e^{i\varphi}\left[-\left(x-iy\right)\hat{z}+kxy\left(\hat{x}+i\hat{y}\right)\right]
\end{eqnarray}

First, it is easy to check that the \textbf{Gauss's law} in vacuum is perfectly fulfilled, \ie 
\begin{equation}
    \nabla\cdot\mathbf{E}= B_{1}e^{i\varphi}\omega\left[-\left(x-iy\right)ik+ky+ikx\right]=0.
\end{equation}
The \textbf{Magnetic Gauss' law} for the considered field reads
\begin{equation}
    \nabla\cdot\mathbf{B}= B_{1}e^{i\varphi}ik^{2}\left(y+ix\right).
\end{equation}
The equation is not fully satisfied, but the residuals are of order $O\left(k^{2}r\right)$, so the possible correction term to the magnetic field is of the order of  $\left(kr\right)^{2}$ because integrating over the divergence operator $\nabla\cdot\mathbf{B}$ adds the length scale of the order of $r$ to the magnetic field.

Similarly, the correction term for the electric field to first non-vanishing order to \textbf{Faraday's law} is of the order of $\left(kr\right)^{2}$ since
\begin{equation}
    \nabla\times\mathbf{E}+ \frac{\partial\mathbf{B}}{\partial t}=B_{1}e^{i\varphi}\omega k^{2}xy\left(\hat{x}+i\hat{y}\right)
\end{equation}

Finally, the non-vanishing terms in the \textbf{Ampere's law} for the considered fields are

\begin{equation}
    \nabla\times\mathbf{B} - \frac{1}{c^{2}}\frac{\partial\mathbf{E}}{\partial t} =\frac{i\omega}{c^{2}}\mathbf{E} =iB_{1}e^{i\varphi}\frac{\omega^{2}}{c^{2}}\left[-\left(x-iy\right)\hat{z}+kxy\left(\hat{x}+i\hat{y}\right)\right]
\end{equation}
which are of second order terms in $\left(\omega r/c\right)$ and $\left(kr\right)$.
This concludes the demonstration that Eqs.~(\ref{eq:TRMF_B})-(\ref{eq:TRMF_E}) for the TRMF scenario are physically consistent to the first order in $\left(\omega r/c\right)$ and $\left(kr\right)$ near the mirror axis.
It is noted that by switching between the magnetic and electric fields, \ie Eqs~(\ref{eq:TREF_E})-(\ref{eq:TREF_B}) and repeating similar calculations as above, one concludes that also the considered TREF electric and magnetic fields are consistent with the Maxwell's equations to the first order in $\left(\omega r/c\right)$ and $\left(kr\right)$.

\end{document}